\documentclass[nofootinbib,amsmath,amssymb,aps,prd,balancelastpage,superscriptaddress]{revtex4-2}
\usepackage[T1]{fontenc}
\usepackage[utf8]{inputenc}
\usepackage{amsmath}
\usepackage{amssymb}
\usepackage{graphicx}

\makeatletter
\usepackage{amsfonts}
\usepackage{array}
\usepackage{multirow}
\usepackage{latexsym}
\usepackage{epsfig}
\usepackage{float}
\usepackage{dsfont}

\usepackage{ragged2e}
\justifying\let\raggedright\justifying
\usepackage{caption}
\newcommand{\be}{\begin{equation}}
\newcommand{\ee}{\end{equation}}

\usepackage[colorlinks]{hyperref}

\makeatother

\begin{document}

%\title{Revisiting Axion-Photon Mixing via Extragalactic Background Light: \\Plateau Regimes, Resonances, and Non-Gaussian Boosts}

\title{The Axion-Photon Mixing and  the Extragalactic Magnetic Background: \\Plateau Regimes, Resonances, and Non-Gaussian Boosts}

\author{{\normalsize Andrea Addazi}}
\affiliation{Center for Theoretical Physics, College of Physics, Sichuan University, China}
%\affiliation{Institute of Astronomy and Astrophysics, School of Physics and Astronomy and Space Sciences, Anqing Normal University, Anqing 246133, China}
\affiliation{INFN, Laboratori Nazionali di Frascati, Via E. Fermi 54, I-00044 Roma, Italy}

\author{{\normalsize Yi-Fu Cai}}
%\email{yifucai@ustc.edu.cn}
\affiliation{Department of Astronomy, School of Physical Sciences,
University of Science and Technology of China, Hefei, Anhui 230026, China}
\affiliation{CAS Key Laboratory for Researches in Galaxies and Cosmology,
School of Astronomy and Space Science, University of Science and Technology of China, Hefei, Anhui 230026, China}

\author{{\normalsize Salvatore Capozziello}}
\affiliation{Dipartimento di Fisica ”E. Pancini”, Università di Napoli “Federico II”, Compl. Univ. di Monte S. Angelo, Edificio G, Via Cinthia, I-80126, Napoli, Italy}
\affiliation{Scuola Superiore Meridionale, Via Mezzocannone 4, I-80134, Napoli, Italy}
\affiliation{Istituto Nazionale di Fisica Nucleare, Sezione di Napoli, Napoli, Italy}

\author{{\normalsize Qingyu Gan}}
\affiliation{Scuola Superiore Meridionale, Via Mezzocannone 4, I-80134, Napoli, Italy}
\affiliation{Istituto Nazionale di Fisica Nucleare, Sezione di Napoli, Napoli, Italy}

\author{{\normalsize Gaetano Lambiase}}
\affiliation{Dipartimento di Fisica ``E.R. Caianiello'', Università di Salerno, I-84084 Fisciano (Sa), Italy}
\affiliation{INFN- Gruppo Collegato di Salerno, I-84084 Fisciano (Sa), Italy}

\begin{abstract}
We present an analytical treatment of Axion-Like-Particle (ALP)---photon mixing with extragalactic background light (EBL) attenuation for constant, Gaussian-stochastic, and non-Gaussian magnetic field configurations—with direct implications for Very High Energy (VHE) gamma-ray observations such as LHAASO, HAWC, and CTA experiments. 
For constant fields, we derive exact probabilities and identify a perturbative plateau regime where photon survival scales as quartic order of magnetic field, isolating the four-point magnetic correlation as a sensitive probe of non-Gaussianity. 
For Gaussian stochastic fields, we obtain—for the first time—analytical formulas for non-exponential-decay components in the strong-attenuation regime. Contrary to the widely used domain-like model, photon survival is suppressed by 4-6 orders of magnitude, while both conversion and survival probabilities exhibit distinct multi-peak structures from mass-equal resonance, stochastic resonance, and EBL attenuation. 
Extending to non-Gaussian fields, we show that non-Gaussianity can enhance photon survival by several orders of magnitude relative to the Gaussian case, potentially explaining the unexpectedly VHE photon event observed by LHAASO. 
Our results demonstrate that stochastic magnetic fields cannot be reduced to domain-like coherence without losing essential physics, and that VHE gamma-ray spectra encode observable information about both the power spectrum and non-Gaussian structure of intergalactic magnetic fields—critical as next-generation observatories push toward PeV sensitivities. 
\end{abstract}

\maketitle

\section{Introduction}

Recently the Large High Altitude Air Shower Observatory (LHAASO) reported
the detection of more than $140$ photons with energies above $3$~TeV, and
$9$ photons above $10$~TeV with high statistical significance from Gamma
Ray Burst 221009A at redshift $0.15$ \cite{LHAASO:2023kyg, LHAASA:2023pay}. 
In addition, the Carpet experiment has reported a photon-like event with an energy of approximately $300$~TeV \cite{Carpet-3Group:2025fcs}. 
Within the framework of conventional physics, very-high-energy photons above the TeV scale originating from distant sources such as gamma-ray bursts are expected to be strongly attenuated due to pair production with extragalactic background light (EBL), i.e., \(\gamma^{\textrm{VHE}}\gamma^{\textrm{EBL}}\rightarrow e^{+}e^{-}\), making their observation challenging. 
One proposed new-physics explanation involves axion-like particles (ALPs), which can oscillate into photons in the presence of a magnetic field, thereby reducing the attenuation of gamma-ray flux by the EBL \cite{Galanti:2022chk, Galanti:2025gaz}. 
Consequently, ALP-photon mixing in the presence of the EBL has been extensively investigated across a range of astrophysical and theoretical contexts. 
These studies encompass model-building efforts \cite{DeAngelis:2011id, Csaki:2003ef, Mirizzi:2009aj, Galanti:2018myb}, observational phenomena in AGNs and blazars \cite{Horns:2012kw, DeAngelis:2007dqd, Li:2024jlc, Meyer:2014epa}, the interpretation of X-ray excesses~\cite{Evoli:2016zhj}, scenarios involving dark photons \cite{Bi:2020ths}, photon polarization effects \cite{Yao:2022col, Cao:2023kdu}, and combined analyses with neutrino emissions \cite{Dekker:2025vcg, Eckner:2022rwf}.

In the studies of ALP-photon mixing—both with and without the influence of EBL—the treatment of magnetic fields constitutes a critical aspect. Methodologically, existing approaches can be broadly divided into discretizations and integral methods. 
The discretization approach is typically implemented via domain-like models, wherein the magnetic field is assumed to be constant within each domain, allowing for an exact solution to the ALP-photon mixing equations \cite{Mirizzi:2009aj, DeAngelis:2011id, Horns:2012kw, Meyer:2014epa, Evoli:2016zhj, Galanti:2018myb}. 
In contrast, the integral approach is based on perturbative treatments that integrate over spatially varying magnetic field configurations \cite{carenza2021turbulent, Mirizzi:2007hr, Marsh:2021ajy, Addazi:2024mii}. 
On the other hand, ALP-photon mixing can be categorized based on the nature of the magnetic field configuration. The first type involves deterministic configurations, which are typically associated with specific astrophysical environments such as the magnetosphere of neutron stars \cite{McDonald:2024nxj} or the regular large-scale magnetic fields in galaxy clusters and galaxies~\cite{Galanti:2018nvl, Troitsky:2023uwu}. 
The second type concerns stochastic magnetic field configurations, which are more prevalent in cosmic settings, including turbulent fields in clusters, galaxies, and jets, as well as primordial magnetic fields and the magnetized cosmic web. To characterize stochasticity statistically, correlation functions are commonly employed. 
The phenomenon of ALP-photon mixing in stochastic magnetic fields has been studied extensively using several forms of two-point correlation functions. 
These include monochromatic and scale-invariant spectra \cite{Addazi:2024mii}, turbulent spectra~\cite{Mirizzi:2007hr, Meyer:2014epa, Kachelriess:2021rzc, Horns:2012kw, Angus:2013sua}, as well as more general spectral shapes \cite{Marsh:2021ajy}. 
Concerning higher order correlation functions, effects of the magnetic field with Gaussian and non-Gaussian distributions on ALP-photon mixing have been explored in Refs.~\cite{Carenza:2022zmq, Montanino:2017ara, carenza2021turbulent}. 

In this work, we revisit ALP-photon mixing in the presence of EBL attenuation for both constant and stochastic magnetic field configurations. 
For the constant field case, we derive the exact solution which generalizes
previous results in Refs.~\cite{Csaki:2003ef, DeAngelis:2011id, Cao:2023kdu}. 
For a stochastic field with a Gaussian distribution, we adopt the integral
approach and, for the first time, obtain the analytical formula of the non-exponential-decay term in the ALP-photon conversion and photon survival probabilities in the strong EBL region. We clarify the perturbative conditions for both constant and stochastic magnetic fields and find that they are satisfied throughout most of the physically relevant parameter space. 
Interestingly, the double oscillation process \(\gamma\rightarrow a\rightarrow\gamma\) implies that, within the perturbative regime, the photon survival probability is dominated by terms at quartic order in the magnetic field strength, i.e., \(\mathcal{O}(B^{4})\). This provides a natural framework for investigating the impact of non-Gaussian magnetic fields on ALP-photon mixing. 

The paper is organized as follows. In Section~\ref{sec:Constant-magnetic-field}, we solve the ALP-photon mixing equations in a constant magnetic field. Section~\ref{sec:Stochastic-magnetic-field} extends this analysis to Gaussian stochastic magnetic fields, where we apply the solution to realistic physical scenarios and compare the predictions of the domain-like model with those of the integral approach. 
In Section~\ref{sec:Non-Gaussian-magnetic-field}, we investigate the photon survival probability in the presence of non-Gaussian magnetic fields. Finally, we present a discussion of our findings in Section~\ref{sec:Discussions and Remarks} and conclude in Section~\ref{sec:Conclusions}.

\section{Constant magnetic field}
\label{sec:Constant-magnetic-field}

An axion-like particle (ALP) field \(a\), characterized by mass \(m_a\) and two-photon coupling \(g_{a\gamma}\), can oscillate into a photon (i.e., the electromagnetic potential \(\mathcal{A}_\mu\)) in the presence of a transverse magnetic background field. 
For relativistic ALPs with frequency \(\omega\) satisfying \(m_a \ll \omega\), this oscillation gives rise to an ALP-photon mixing system whose dynamics are governed by the following equations of motion \cite{raffelt1988mixing, Galanti:2018myb}:
\begin{eqnarray}
\partial_{l}\left(\begin{array}{c}
\mathcal{A}_{x}\\
\mathcal{A}_{y}\\
a
\end{array}\right) & = & iK\left(\begin{array}{c}
\mathcal{A}_{x}\\
\mathcal{A}_{y}\\
a
\end{array}\right)=-i\left(\begin{array}{ccc}
\omega+\Delta_{xx} & \Delta_{R} & 0\\
\Delta_{R} & \omega+\Delta_{yy} & \Delta_{B}\\
0 & \Delta_{B} & \omega+\Delta_{a}
\end{array}\right)\left(\begin{array}{c}
\mathcal{A}_{x}\\
\mathcal{A}_{y}\\
a
\end{array}\right),\label{eq:EoM-const}
\end{eqnarray}
where $l$ the propagating direction. Given a homogeneous transverse magnetic field, one may rotate the coordinate frame so that the field aligns with the \(y\)-axis. In this frame, the relevant mixing term is given by \(\Delta_{B}=g_{a\gamma}B/2\). The $\Delta_{a}\equiv \Delta_{a}(\omega)=-m_{a}^{2}/2\omega$ term corresponds to the effective ALP mass contribution. 

Regarding the photon terms, firstly, the Faraday rotation term $\Delta_{R}$ 
couples the two photon polarizations. 
While it is relevant for the analysis of polarized photon sources and irrelevant to this study, we therefore set $\Delta_{R}=0$. 
Secondly, $\Delta_{xx}$ and $\Delta_{yy}$ include both real and imaginary parts, which we assume to be equal for simplicity and denote as $\Delta_{xx}=\Delta_{yy}=\Delta_{\gamma}=A-ib$. 
The real part $A$ contains conventional contributions such as plasma effect, QED corrections and CMB effects (see Eq.~\ref{eq:mass-terms}). 
For very high energy photons propagating through the intergalactic medium, it undergoes pair production interacting with the optical and infrared EBL photons, i.e. $\gamma^{\textrm{VHE}}\gamma^{\textrm{EBL}}\rightarrow e^{+}e^{-}$ \cite{Franceschini:2017iwq}. 
This leads to an EBL attenuation effect in the ALP-photon mixing, which can be described by adding an imaginary part $b=1/\left(2\lambda_{\gamma}\right)$ with the photon mean free path $\lambda_{\gamma}(\omega)$. 
In this work, we adopt the semi-analytical expression used in \cite{Mirizzi:2009aj, Montanino:2017ara}, 
\begin{eqnarray}
 \lambda_{\gamma} & = & 1.2\times10^{3}\left(\frac{\omega}{\textrm{TeV}}\right)^{-1.55}\textrm{Mpc} .
\end{eqnarray}
The front factor $1.2$ is chosen such that $\lambda_{\gamma}=\textrm{Mpc}$
at $\omega=100\textrm{TeV}$, serving as a practical benchmark point.For later convenience, we define the oscillation length of ALP-photon
mixing as $l_{osc}=2/\sqrt{A^{2}+\Delta_{B}^{2}}$ in constant magnetic
field and $l_{osc}=2/|A|$ in stochastic magnetic field. 

The dynamics of ALP-photon mixing can be solved in terms of the evolution
operator $U(l)$. The evolution operator is actually a $3\times 3$ matrix  defined as 
$\left(A_{\perp}(l),A_{\|}(l),a(l)\right)^{T}=U(l)\left(A_{\perp}(0),A_{\|}(0),a(0)\right)^{T}$,
where we set $l_{0}=0$ as the initial point. 
The EoM for $U(l)$ reads as
\begin{eqnarray}
\partial_{l}U(l) & = & iKU(l),\label{eq:EoM-U}
\end{eqnarray}
with the initial condition $U(0)=I_{3\times3}$. Because $K$ is $l$-independent, the explicit closed solution is simply $U(l)=e^{iKl}$, and non-vanishing components are given by 
\begin{eqnarray}
U_{11} & = & e^{-i\left(\omega+\Delta_{\gamma}\right)l},\nonumber \\
U_{22} & = & \frac{1}{2s}e^{-\frac{1}{2}i\left(2\omega+\Delta_{\gamma}+\Delta_{a}\right)l}\left(\left(\Delta_{a}-\Delta_{\gamma}+s\right)e^{\frac{1}{2}isl}-\left(\Delta_{a}-\Delta_{\gamma}-s\right)e^{-\frac{1}{2}isl}\right),\nonumber \\
U_{23} & = & U_{32}=-\frac{\Delta_{B}}{s}e^{-\frac{1}{2}i\left(2\omega+\Delta_{\gamma}+\Delta_{a}\right)l}\left(e^{\frac{1}{2}isl}-e^{-\frac{1}{2}isl}\right),\nonumber \\
U_{33} & = & \frac{1}{2s}e^{-\frac{1}{2}i\left(2\omega+\Delta_{\gamma}+\Delta_{a}\right)l}\left(\left(\Delta_{a}-\Delta_{\gamma}+s\right)e^{-\frac{1}{2}isl}-\left(\Delta_{a}-\Delta_{\gamma}-s\right)e^{\frac{1}{2}isl}\right),\label{eq:constB-U}
\end{eqnarray}
where $s=\sqrt{4\Delta_{B}^{2}+\left(\Delta_{\gamma}-\Delta_{a}\right)^{2}}$.
All other components of $U$ vanish. The relevant probabilities, namely
ALP-photon conversion probability, photon-ALP conversion probability,
photon survival probability and ALP survival probability, are related
to $U$ by (see Appendix A) 
\begin{eqnarray}
\mathcal{P}_{a\rightarrow\gamma} & = & U_{13}U_{13}^{*}+U_{23}U_{23}^{*},\nonumber \\
\mathcal{P}_{\gamma\rightarrow a} & = & \frac{1}{2}\left(U_{31}U_{31}^{*}+U_{32}U_{32}^{*}\right),\nonumber \\
\mathcal{P}_{\gamma\rightarrow\gamma} & = & \frac{1}{2}\left(U_{11}U_{11}^{*}+U_{12}U_{12}^{*}+U_{21}U_{21}^{*}+U_{22}U_{22}^{*}\right),\nonumber \\
\mathcal{P}_{a\rightarrow a} & = & U_{33}U_{33}^{*}.\label{eq:P-U}
\end{eqnarray}
The square root $s$ can be evaluated as
\begin{eqnarray}
s & = & \begin{cases}
\alpha+i\beta, & \hfill\mathcal{I}\geq0,\\
\alpha-i\beta, & \hfill\mathcal{I}<0,
\end{cases}\label{eq:RootIm}
\end{eqnarray}
with 
\begin{eqnarray}
\alpha=\frac{\sqrt{\sqrt{\mathcal{R}^{2}+\mathcal{I}^{2}}+\mathcal{R}}}{\sqrt{2}}, & \quad & \beta=\frac{\sqrt{\sqrt{\mathcal{R}^{2}+\mathcal{I}^{2}}-\mathcal{R}}}{\sqrt{2}},\quad\mathcal{R}=4\Delta_{B}^{2}+A^{2}-b^{2},\quad\mathcal{I}=-2bA.
\end{eqnarray}
Combining Eqs. (\ref{eq:constB-U}), (\ref{eq:P-U}) and (\ref{eq:RootIm}),
we obtain the relevant probabilities after traveling a distance $d$,
\begin{eqnarray}
\mathcal{P}_{\gamma\rightarrow a}\left(d\right) & = & \frac{1}{2}P_{a\rightarrow\gamma}=\frac{1}{2}\frac{\Delta_{B}^{2}}{\alpha^{2}+\beta^{2}}e^{-bd}\left\{ 2\textrm{cosh}\left(d\beta\right)-2\textrm{cos}\left(d\alpha\right)\right\} ,\nonumber \\
\mathcal{P}_{\gamma\rightarrow\gamma}\left(d\right) & = & \frac{1}{2}e^{-2bd}+\frac{1}{4(\alpha^{2}+\beta^{2})}e^{-bd}\left\{ \left(A^{2}+b^{2}+\alpha^{2}+\beta{}^{2}\right)\textrm{cosh}\left(d\beta\right)-2\left(\alpha\left|A\right|+\beta b\right)\textrm{sinh}\left(d\beta\right)\right.\nonumber \\
 &  & \left.-\left(A^{2}+b^{2}-\alpha^{2}-\beta{}^{2}\right)\textrm{cos}\left(d\alpha\right)+2\left(\beta\left|A\right|-\alpha b\right)\textrm{sin}\left(d\alpha\right)\right\} ,\nonumber \\
\mathcal{P}_{a\rightarrow a}\left(d\right) & = & \frac{1}{2(\alpha^{2}+\beta^{2})}e^{-bd}\left\{ \left(A^{2}+b^{2}+\alpha^{2}+\beta{}^{2}\right)\textrm{cosh}\left(d\beta\right)+2\left(\alpha\left|A\right|+\beta b\right)\textrm{sinh}\left(d\beta\right)\right.\nonumber \\
 &  & \left.-\left(A^{2}+b^{2}-\alpha^{2}-\beta{}^{2}\right)\textrm{cos}\left(d\alpha\right)-2\left(\beta\left|A\right|-\alpha b\right)\textrm{sin}\left(d\alpha\right)\right\} .\label{eq:const-Ps}
\end{eqnarray}
These expressions are valid for both $\mathcal{I}\geq0$ and $\mathcal{I}<0$.
In the limit \(b \rightarrow 0\), they reduce to the conventional ALP-photon mixing probabilities in the absence of EBL effects. In the limit \(A \rightarrow 0\), they recover the results of Refs.~\cite{Csaki:2003ef,Cao:2023kdu}. 

\begin{figure}
\includegraphics[scale=0.7]{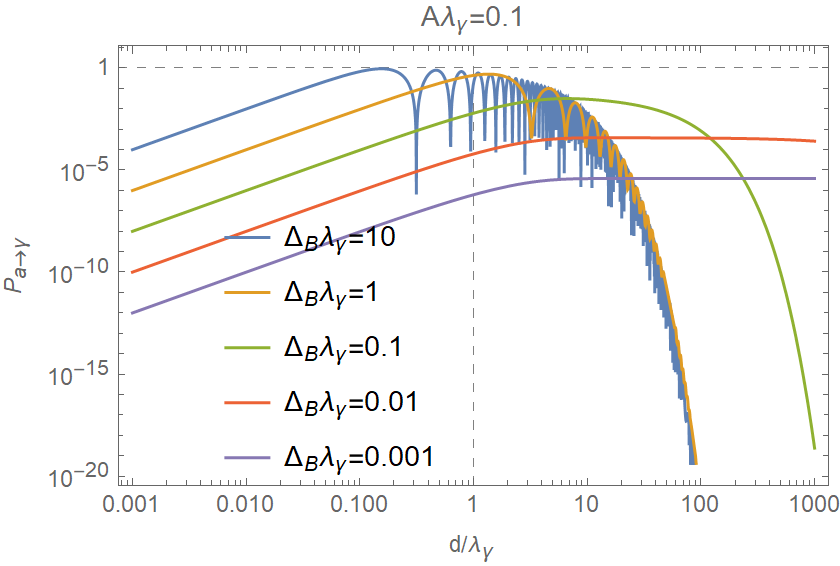}\includegraphics[scale=0.7]{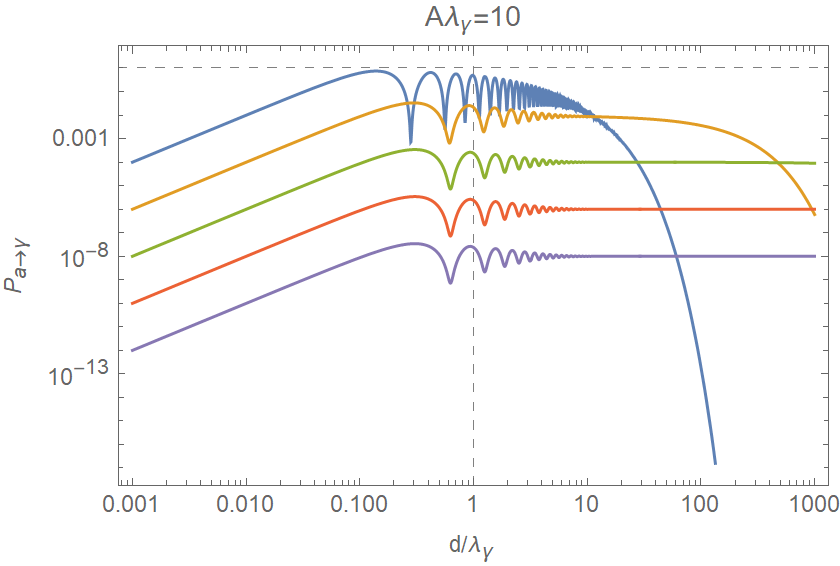}

\includegraphics[scale=0.7]{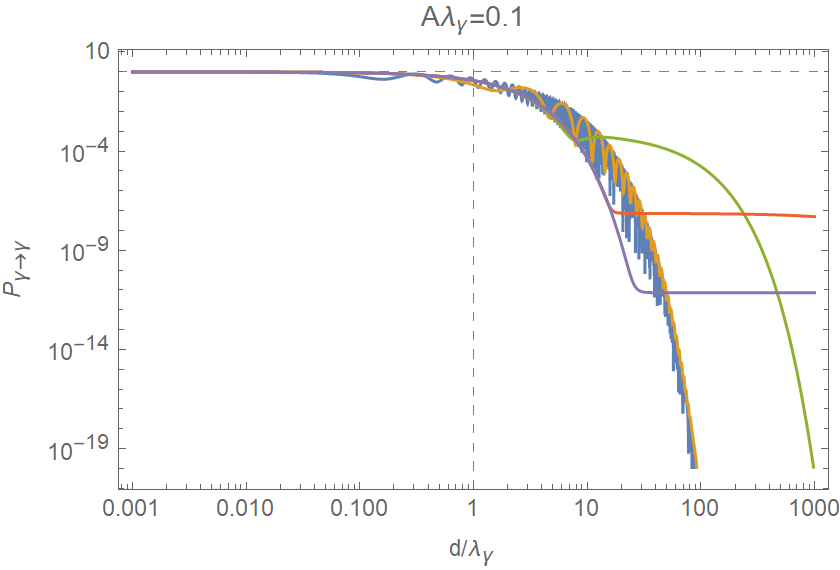}\includegraphics[scale=0.7]{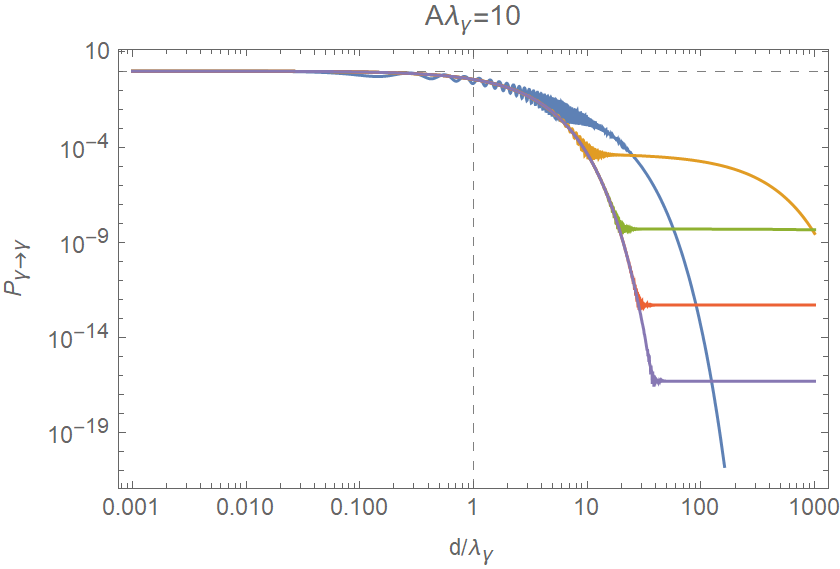}

\caption{The ALP-photon conversion probability \(\mathcal{P}_{a\rightarrow\gamma}\) (upper panels) and the photon survival probability \(\mathcal{P}_{\gamma\rightarrow\gamma}\) (lower panels) in a constant magnetic field for various parameter sets. All dimensional quantities are normalized to \(\lambda_{\gamma}\). In the regime \(d \ll \lambda_{\gamma}\), the probabilities reproduce the conventional ALP-photon mixing results in the absence of EBL attenuation. For \(\lambda_{\gamma} \ll d \ll \left(1 + A^{2}\lambda_{\gamma}^{2}\right) / \left(\Delta_{B}^{2}\lambda_{\gamma}\right)\), both probabilities lie within the perturbative regime and remain approximately constant, exhibiting power-law scaling with the magnetic field as \(\mathcal{P}_{a\rightarrow\gamma} \sim B^{2}\) and \(\mathcal{P}_{\gamma\rightarrow\gamma} \sim B^{4}\). At larger distances, the probabilities decay exponentially due to EBL absorption.}

\label{fig:P-const}
\end{figure}

Using Eq.~\eqref{eq:const-Ps}, we compute the probabilities for several representative parameter sets, with the results shown in Fig.~\ref{fig:P-const}. For certain parameter choices, both the ALP-photon conversion and photon survival probabilities remain approximately constant in the regime \(d \gg \lambda_{\gamma}\), which corresponds to the perturbative regime of the mixing. To clarify this behavior, we expand Eq.~\eqref{eq:const-Ps} in powers of \(g_{a\gamma}B\lambda_{\gamma}\) and focus on the large-distance limit \(d \gg \lambda_{\gamma}\), where exponential-decay terms are strongly suppressed. Then we obtain (see
Appendix B)
\begin{eqnarray}
\mathcal{P}_{\gamma\leftrightarrow a} & \sim & \left(g_{a\gamma}B\lambda_{\gamma}\right)^{2}\frac{1}{1+A^{2}\lambda_{\gamma}^{2}}+\left(g_{a\gamma}B\lambda_{\gamma}\right)^{4}\frac{d}{\lambda_{\gamma}}\frac{1}{1+A^{4}\lambda_{\gamma}^{4}}+\left(g_{a\gamma}B\lambda_{\gamma}\right)^{6}\frac{d^{2}}{\lambda_{\gamma}^{2}}\frac{1}{1+A^{6}\lambda_{\gamma}^{6}}+...,\nonumber \\
\mathcal{P}_{\gamma\rightarrow\gamma} & \sim & \left(g_{a\gamma}B\lambda_{\gamma}\right)^{4}\frac{1}{1+A^{4}\lambda_{\gamma}^{4}}+\left(g_{a\gamma}B\lambda_{\gamma}\right)^{6}\frac{d}{\lambda_{\gamma}}\frac{1}{1+A^{6}\lambda_{\gamma}^{6}}+\left(g_{a\gamma}B\lambda_{\gamma}\right)^{8}\frac{d^{2}}{\lambda_{\gamma}^{2}}\frac{1}{1+A^{8}\lambda_{\gamma}^{8}}+...,\label{eq:P-expansion}
\end{eqnarray}
Neglecting higher-order terms requires the perturbative condition
\begin{equation}
\left(g_{a\gamma}B\lambda_{\gamma}\right)^{2}\frac{d}{\lambda_{\gamma}}\frac{1}{1+A^{2}\lambda_{\gamma}^{2}} \ll 1.
 \end{equation}
When this condition is satisfied, both the leading-order term of \(\mathcal{P}_{\gamma\leftrightarrow a}\) (at order \(B^{2}\)) and that of \(\mathcal{P}_{\gamma\rightarrow\gamma}\) (at order \(B^{4}\)) become independent of the distance \(d\). Indeed, as shown in Fig.
(\ref{fig:P-const}), for example with $\Delta_{B}\lambda_{\gamma}=0.1$,
the probabilities $\mathcal{P}_{\gamma\leftrightarrow a}$ and $\mathcal{P}_{\gamma\rightarrow\gamma}$
begin to decay exponentially at $d/\lambda_{\gamma}>100$ for $A\lambda_{\gamma}=0.1$,
and they remain constant $\mathcal{P}_{\gamma\leftrightarrow a}\simeq10^{-4}$
and $\mathcal{P}_{\gamma\rightarrow\gamma}\simeq10^{-8}$ at $1\ll d/\lambda_{\gamma}<1000$
for $A\lambda_{\gamma}=10$. A similar distinct plateau profile of photon
survival probability has been observed in Refs. \cite{Csaki:2003ef,Horns:2012kw,Cao:2023kdu}.
As clearly shown in Eq. (\ref{eq:P-expansion}) and Fig. (\ref{fig:P-const}),
within the perturbative region and at $d\gg\lambda_{\gamma}$, there
exists a relation $\mathcal{P}_{\gamma\rightarrow\gamma}\sim\mathcal{P}_{\gamma\leftrightarrow a}^{2}$,
which can be interpreted as a double conversion $\gamma\rightarrow a\rightarrow\gamma$.
However, no simply relation exists between $P_{\gamma\leftrightarrow a}$
and $P_{\gamma\rightarrow\gamma}$ beyond the perturbative regime.
Moreover, we recall that in the absence (or for negligible) EBL effects, the perturbative regime of ALP-photon mixing corresponds to weak mixing, characterized by the condition \(g_{a\gamma}B l_{\mathrm{osc}} \ll 1\)~\cite{raffelt1988mixing}. In contrast, in the presence of strong EBL, the relevant perturbative parameter becomes \(g_{a\gamma}B \lambda_{\gamma} \ll 1\), as justified by the expansion in Eq.~\eqref{eq:P-expansion} and the plateau behavior observed in Fig.~\ref{fig:P-const}.

In general, the magnetic field is not coherent across the entire propagation path. To account for this, a domain-like approach has been developed to model the decoherence of the magnetic field. In this framework, the total space is partitioned into \(N\) domains, each with a size equal to the typical correlation length \(\lambda_B\) of the varying magnetic field. Within each domain, the magnetic field vector is assumed to have a fixed but randomly oriented direction. For simplicity, we adopt a reduced version of the domain-like model in which only the \(y\)-component of the magnetic field is retained, implemented via vector projection.
Specifically,
we replace Eq. (\ref{eq:EoM-const}) with $\Delta_{B}\rightarrow\Delta_{B}\textrm{cos}\vartheta$,
where $\vartheta$ is a random variable uniformly distributed within $\left[0,\pi\right]$.
Let $F_{\gamma}^{(0)}$ and $F_{a}^{(0)}$ denote the initial photon
and ALP fluxes in the first domain. The iterative relation of photon
and ALP fluxes between patch $n$ and patch $n+1$ is
\begin{eqnarray}
F_{\gamma}^{(n+1)} & = & F_{\gamma}^{(n)}\mathcal{P}_{\gamma\rightarrow\gamma}^{(n)}+F_{a}^{(n)}\mathcal{P}_{a\rightarrow\gamma}^{(n)},\nonumber \\
F_{a}^{(n+1)} & = & F_{\gamma}^{(n)}\mathcal{P}_{\gamma\rightarrow a}^{(n)}+F_{a}^{(n)}\mathcal{P}_{a\rightarrow a}^{(n)},\label{eq:Simplied-DL}
\end{eqnarray}
where $\mathcal{P}^{(n)}$ denote the probability in $n$-th patch.
For an initial state consisting purely of ALP (photon), the overall
ALP-photon conversion probability (photon survival probability) after
traversing $N$ patches is given by $\mathcal{P}_{a\rightarrow\gamma}^{cell}=F_{\gamma}^{(N)}/F_{a}^{(0)}$
( $\mathcal{P}_{\gamma\rightarrow\gamma}^{cell}=F_{\gamma}^{(N)}/F_{\gamma}^{(0)}$).
Later we will compare the ALP-photon mixing probabilities in stochastic
magnetic field using domain-like approach with those obtained using
the integral approach in next section.

\section{Stochastic magnetic field}

\label{sec:Stochastic-magnetic-field}

The perturbative expansion of the probabilities in the constant magnetic field case, given in Eq.~\eqref{eq:P-expansion}, motivates a similar treatment for stochastic magnetic field configurations. In this scenario, one must expand the solution up to \(\mathcal{O}(B^{2})\) for \(P_{a\rightarrow\gamma}\) and up to \(\mathcal{O}(B^{4})\) for \(P_{\gamma\rightarrow\gamma}\) (see Appendix B). To implement this, we first replace the constant mixing term \(\Delta_{B}\) in Eq.~\eqref{eq:EoM-const} with a spatially varying counterpart and express the kernel matrix as
\begin{eqnarray}
K & = & \left(\begin{array}{ccc}
\omega+\Delta_{\gamma}\qquad & 0\qquad & \Delta_{Bx}(l)\\
0\qquad & \omega+\Delta_{\gamma}\qquad & \Delta_{By}(l)\\
\Delta_{Bx}(l)\qquad & \Delta_{By}(l)\qquad & \omega+\Delta_{a}
\end{array}\right),
\end{eqnarray}
where $\Delta_{Bx}(l)=\frac{1}{2}g_{a\gamma}B_{x}(l)$ and $\Delta_{By}(l)=\frac{1}{2}g_{a\gamma}B_{y}(l)$.
The EoM and initial condition of the evolution operator remain Eq.
\ref{eq:EoM-U}, i.e. $\partial_{l}U\left(l\right)=iK(l)U\left(l\right)$
and $U\left(l_{0}=0\right)=I_{3\times3}$. Due to the $l$-dependent $K(l)$,
the solution can no longer be expressed in closed analytic form
as in the constant magnetic field case. Consequently, we employ the perturbative framework introduced in Refs.~\cite{raffelt1988mixing,Mirizzi:2007hr,Addazi:2024kbq,Addazi:2024mii}.
The kernel is split as $K(l)=K_{0}+\delta K(l)$, where $K_{0}$ and
$\delta K(l)$ contain the diagonal and off-diagonal components, respectively.
Solving the EoM iteratively yields a Dyson perturbation series {\small
\begin{eqnarray}
U\left(l\right) & = & e^{ilK_{0}}+e^{ilK_{0}}i\int_{0}^{l}dl_{1}e^{-il_{1}K_{0}}\delta K(l_{1})e^{il_{1}K_{0}}\nonumber \\
 &  & +e^{ilK_{0}}i\int_{0}^{l}dl_{1}e^{-il_{1}K_{0}}\delta K(l_{1})e^{il_{1}K_{0}}i\int_{0}^{l_{1}}dl_{2}e^{-il_{2}K_{0}}\delta K(l_{2})e^{il_{2}K_{0}}\nonumber \\
 &  & +e^{ilK_{0}}i\int_{0}^{l}dl_{1}e^{-il_{1}K_{0}}\delta K(l_{1})e^{il_{1}K_{0}}i\int_{0}^{l_{1}}dl_{2}e^{-il_{2}K_{0}}\delta K(l_{2})e^{il_{2}K_{0}}i\int_{0}^{l_{2}}dl_{3}e^{-il_{3}K_{0}}\delta K(l_{3})e^{il_{3}K_{0}}\nonumber \\
 &  & +e^{ilK_{0}}i\int_{0}^{l}dl_{1}e^{-il_{1}K_{0}}\delta K(l_{1})e^{il_{1}K_{0}}i\int_{0}^{l_{1}}dl_{2}e^{-il_{2}K_{0}}\delta K(l_{2})e^{il_{2}K_{0}}i\int_{0}^{l_{2}}dl_{3}e^{-il_{3}K_{0}}\delta K(l_{3})e^{il_{3}K_{0}}i\int_{0}^{l_{3}}dl_{4}e^{-il_{4}K_{0}}\delta K(l_{4})e^{il_{4}K_{0}}\nonumber \\
 &  & +\mathcal{O}\left(\delta K^{5}\right)\, .\label{eq:U-expan}
\end{eqnarray}
}{\small\par}

In addition, the spatially varying magnetic field is assumed to be
stochastic and non-helical, characterized by the two-point correlation
function 
\begin{eqnarray}
\left\langle \mathbf{B}_{i}(\boldsymbol{x})\mathbf{B}_{j}\left(\boldsymbol{x}'\right)\right\rangle  & = & \frac{1}{(2\pi)^{3}}\int d^{3}\boldsymbol{k} \, e^{i\boldsymbol{k}\cdot\left(\boldsymbol{x}-\boldsymbol{x}'\right)}\left(\delta_{ij}-\frac{\boldsymbol{k}_{i}}{k}\frac{\boldsymbol{k}_{j}}{k}\right)P_{B}(k)\,,
\end{eqnarray}
where $k=|\boldsymbol{k}|$ and $P_{B}(k)$ is the power spectrum.
Note that all $l$-integrals in Eq. (\ref{eq:U-expan}) are independent
from the azimuthal angle $\varphi$; thus we can immediately integrate
on $\varphi$ and obtain 
\begin{eqnarray}
\left\langle B_{x}(l)B_{x}(l')\right\rangle  & = & \left\langle B_{y}(l)B_{y}(l')\right\rangle =\frac{\pi}{(2\pi)^{3}}\int k^{2}dk\int_{-1}^{1}dte^{ikt\left(l-l'\right)}\left(1+t^{2}\right)P_{B}(k)\nonumber \\
\left\langle B_{x}(l)B_{y}(l')\right\rangle  & = & \left\langle B_{y}(l)B_{x}(l')\right\rangle =0\label{eq: Bcorr-xy}
\end{eqnarray}
where $t=\textrm{cos}\theta$ with the $\theta$ the polar angle with
respect to the $l$-direction.

From the perturbative expansion of \(U(l)\) in Eq.~\eqref{eq:U-expan}, we extract the components \(U_{13}\) and \(U_{23}\) and obtain the ALP-photon conversion probability to \(\mathcal{O}(B^{2})\):
\begin{eqnarray}
\mathcal{P}_{a\rightarrow\gamma}(d) & = & \frac{1}{4}g_{a\gamma}^{2}e^{-2bd}\int_{0}^{d}dl_{1}\int_{0}^{d}dl_{2}e^{i(l_{1}-l_{2})A}e^{(l_{1}+l_{2})b}\left(\left\langle B_{x}(l_{1})B_{x}(l_{2})\right\rangle +\left\langle B_{y}(l_{1})B_{y}(l_{2})\right\rangle \right)\nonumber \\
 & = & \frac{1}{4}g_{a\gamma}^{2}\frac{1}{(2\pi)^{3}}\int dk2\pi k^{2}P_{B}\int_{-1}^{1}dt\left(1+t^{2}\right)\frac{1+e^{-2bd}-e^{-bd}2\textrm{cos}\left(\left(A+kt\right)d\right)}{\left(A+kt\right)^{2}+b^{2}}.\label{eq:stoB-P-a-ph}
\end{eqnarray}
The complete expression after integrating over $t$ is presented in
Eq. (\ref{eq:appen-P-aph-sto}) in Appendix C. The first term in the
nominator corresponds to the non exponential-decay component of $\mathcal{P}_{a\rightarrow\gamma}$.
The conversion probability $\gamma-ALP$ is given by $P_{\gamma\rightarrow a}=\frac{1}{2}P_{a\rightarrow\gamma}$.
In the limit $b\rightarrow0$, the result consistently reduces to
the conventional ALP-photon conversion in the stochastic magnetic
field without EBL attenuation \cite{Addazi:2024kbq}.

\begin{figure}
\includegraphics[scale=0.7]{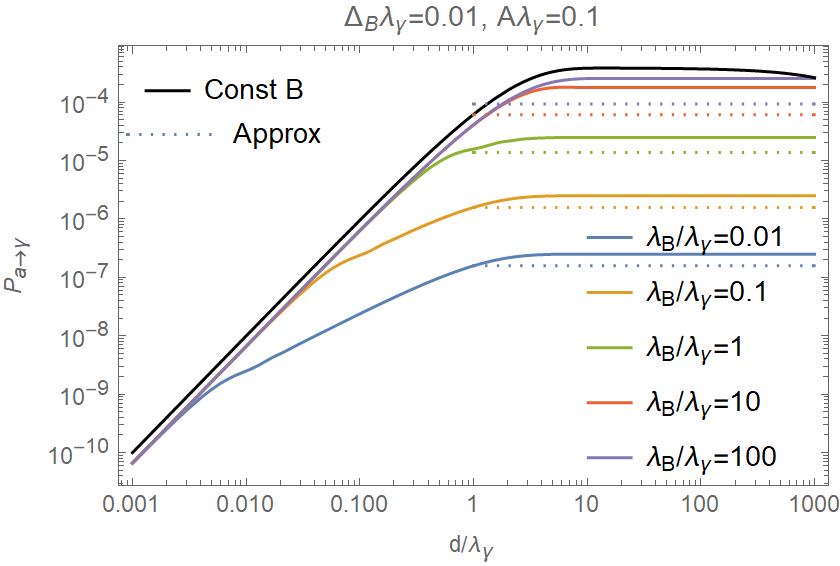}\includegraphics[scale=0.7]{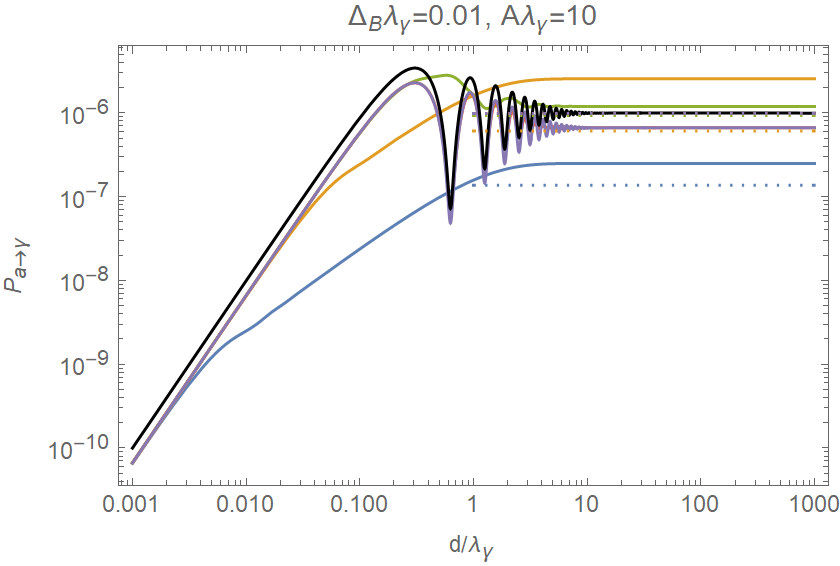}

\includegraphics[scale=0.7]{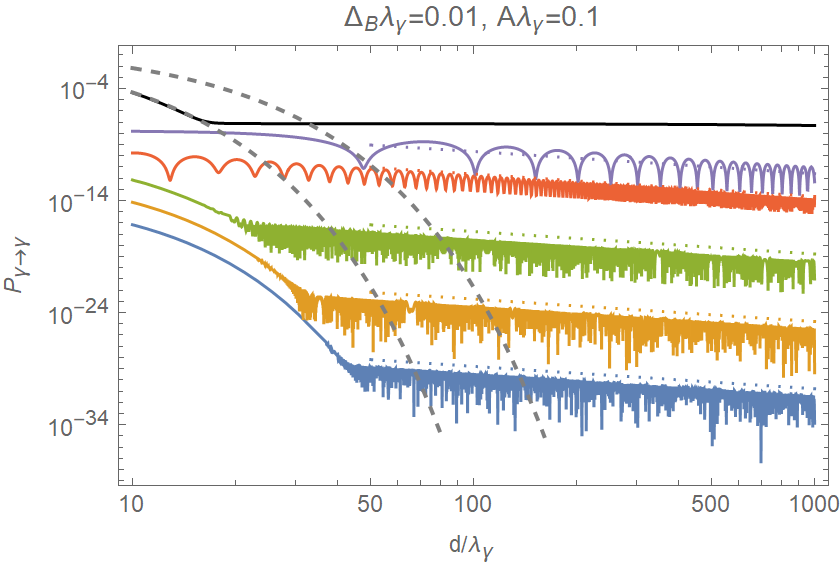}\includegraphics[scale=0.7]{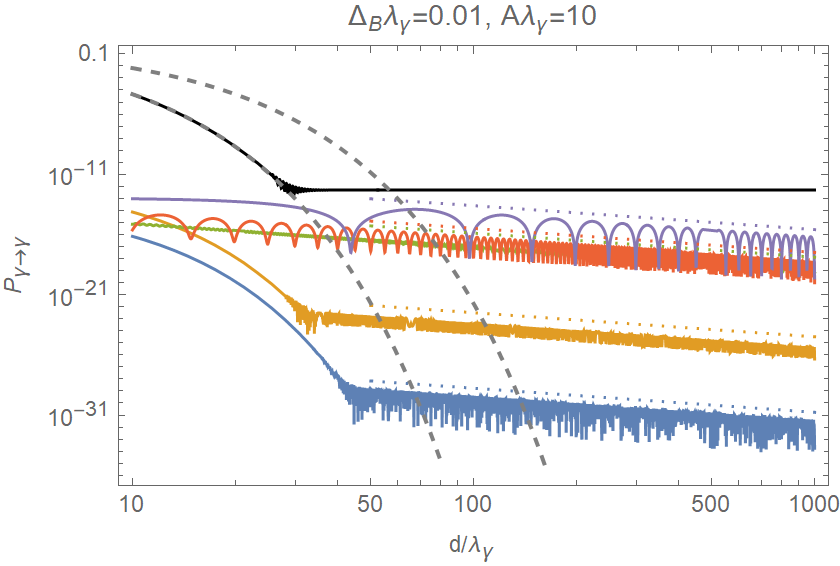}

\caption{The ALP-photon mixing probabilities $\mathcal{P}_{a\rightarrow\gamma}$
and $\mathcal{P}_{\gamma\rightarrow\gamma}$ in the stochastic magnetic
field with Gaussian distribution and the monochromatic spectrum. All
dimensional quantities are properly normalized to $\lambda_{\gamma}$.
Eq.\ref{eq:P-parameterization} provides an approximation 
(colored  dash lines) which that holds in the regime $d/\lambda_{\gamma} \gg 1$. 
{In the first plot illustrating ALP-photon conversion, the initial linear growth is a signature of stochastic resonance (see Section III for further details). }
The gray dashed lines correspond to $e^{-d/\lambda_{\gamma}}$ and
$e^{-d/(2\lambda_{\gamma})}$, above which the exponential-decay component
of the $\mathcal{P}_{\gamma\rightarrow\gamma}$ becomes subdominant
and can be safely neglected compared to the non exponential-decay
contribution.}

\label{fig:P-stoc}
\end{figure}

Turning to the photon survival probability, the perturbative expansion must be extended to quartic order in \(B\), i.e., \(\mathcal{O}(B^{4})\):
\begin{eqnarray}
\mathcal{P}_{\gamma\rightarrow\gamma} & = & \mathcal{P}_{\gamma\rightarrow\gamma}^{(0)}+\mathcal{P}_{\gamma\rightarrow\gamma}^{(1)}+\mathcal{P}_{\gamma\rightarrow\gamma}^{(2)},
\end{eqnarray}
where {\small
\begin{eqnarray}
\mathcal{P}_{\gamma\rightarrow\gamma}^{(0)} & = & e^{-2bd},\nonumber \\
\mathcal{P}_{\gamma\rightarrow\gamma}^{(1)} & = & -\frac{1}{2}\frac{1}{4}g_{a\gamma}^{2}e^{-2bd}\int_{0}^{d}dl_{1}\int_{0}^{l_{1}}dl_{2}e^{\left(l_{1}-l_{2}\right)b}\left(e^{i\left(l_{1}-l_{2}\right)A}+e^{-i\left(l_{1}-l_{2}\right)A}\right)\left(\left\langle B_{x}\left(l_{1}\right)B_{x}\left(l_{2}\right)\right\rangle +\left\langle B_{y}\left(l_{1}\right)B_{y}\left(l_{2}\right)\right\rangle \right),\nonumber \\
\mathcal{P}_{\gamma\rightarrow\gamma}^{(2)} & = & \frac{1}{2}\frac{1}{16}g_{a\gamma}^{4}e^{-2bd}\left\{ \int_{0}^{d}dl_{1}\int_{0}^{l_{1}}dl_{2}\int_{0}^{d}dl_{3}\int_{0}^{l_{3}}dl_{4}e^{i\left(l_{1}-l_{2}\right)A+\left(l_{1}-l_{2}\right)b}e^{-i\left(l_{3}-l_{4}\right)A+\left(l_{3}-l_{4}\right)b}\left(\left\langle B_{x}(l_{1})B_{x}(l_{2})B_{x}(l_{3})B_{x}(l_{4})\right\rangle \right.\right.\nonumber \\
 &  & \left.+\left\langle B_{y}(l_{1})B_{y}(l_{2})B_{y}(l_{3})B_{y}(l_{4})\right\rangle +\left\langle B_{x}(l_{1})B_{y}(l_{2})B_{x}(l_{3})B_{y}(l_{4})\right\rangle +\left\langle B_{y}(l_{1})B_{x}(l_{2})B_{y}(l_{3})B_{x}(l_{4})\right\rangle \right)\nonumber \\
 &  & +\int_{0}^{d}dl_{1}\int_{0}^{l_{1}}dl_{2}\int_{0}^{l_{2}}dl_{3}\int_{0}^{l_{3}}dl_{4}e^{\left(l_{1}-l_{2}+l_{3}-l_{4}\right)b}\left(e^{i\left(l_{1}-l_{2}+l_{3}-l_{4}\right)A}+e^{-i\left(l_{1}-l_{2}+l_{3}-l_{4}\right)A}\right)\left(\left\langle B_{x}(l_{1})B_{x}(l_{2})B_{x}(l_{3})B_{x}(l_{4})\right\rangle \right.\nonumber \\
 &  & \left.\left.+\left\langle B_{x}(l_{1})B_{y}(l_{2})B_{y}(l_{3})B_{x}(l_{4})\right\rangle +\left\langle B_{y}(l_{1})B_{y}(l_{2})B_{y}(l_{3})B_{y}(l_{4})\right\rangle +\left\langle B_{y}(l_{1})B_{x}(l_{2})B_{x}(l_{3})B_{y}(l_{4})\right\rangle \right)\right\} .\label{eq:StoB-P-0-1-2}
\end{eqnarray}
}Similarly to constant magnetic field case, we find that the non exponential-decay
term only appears at forth order $B^{4}$ of probability expansion,
namely $\mathcal{P}_{\gamma\rightarrow\gamma}^{(2)}$ (see Appendix
B). To evaluate the four point correlation function, we further assume
that the stochastic magnetic field is Gaussian. Then we are allowed
to apply the Wick theorem to decompose four point correlation functions
into pairings of two point correlation functions:{\small
\begin{eqnarray}
\left\langle B_{x}(l_{1})B_{x}(l_{2})B_{x}(l_{3})B_{x}(l_{4})\right\rangle  & = & \left\langle B_{y}(l_{1})B_{y}(l_{2})B_{y}(l_{3})B_{y}(l_{4})\right\rangle \nonumber \\
 & = & \left\langle B_{x}(l_{1})B_{x}(l_{2})\right\rangle \left\langle B_{x}(l_{3})B_{x}(l_{4})\right\rangle +\left\langle B_{x}(l_{1})B_{x}(l_{3})\right\rangle \left\langle B_{x}(l_{2})B_{x}(l_{4})\right\rangle +\left\langle B_{x}(l_{1})B_{x}(l_{4})\right\rangle \left\langle B_{x}(l_{2})B_{x}(l_{3})\right\rangle ,\nonumber \\
\left\langle B_{x}(l_{1})B_{y}(l_{2})B_{x}(l_{3})B_{y}(l_{4})\right\rangle  & = & \left\langle B_{x}(l_{1})B_{x}(l_{3})\right\rangle \left\langle B_{x}(l_{2})B_{x}(l_{4})\right\rangle ,\label{eq:4PCF-Wick}
\end{eqnarray}
}where we have used Eq. (\ref{eq: Bcorr-xy}). 

In principle, one could substitute Eqs.~\eqref{eq:4PCF-Wick} and~\eqref{eq: Bcorr-xy} into Eq.~\eqref{eq:StoB-P-0-1-2} and evaluate the full integral. However, the resulting expression is lengthy and tedious. Since our primary interest lies in the non-exponential-decay contribution, we observe that it emerges only from the following integral structure:
$\int_{0}^{d}dl_{1}\int_{0}^{l_{1}}dl_{2}\int_{0}^{d}dl_{3}\int_{0}^{l_{3}}dl_{4}$, and we obtain {\small
\begin{eqnarray}
\mathcal{P}_{\gamma\rightarrow\gamma}^{(4)} & \rightarrow & \frac{1}{2}\frac{1}{16}g_{a\gamma}^{4}e^{-2bd}\int_{0}^{d}dl_{1}\int_{0}^{l_{1}}dl_{2}\int_{0}^{d}dl_{3}\int_{0}^{l_{3}}dl_{4}e^{i\left(l_{1}-l_{2}\right)A+\left(l_{1}-l_{2}\right)b}e^{-i\left(l_{3}-l_{4}\right)A+\left(l_{3}-l_{4}\right)b}\nonumber \\
 &  & \times\left(2\left\langle B_{x}(l_{1})B_{x}(l_{2})\right\rangle \left\langle B_{x}(l_{3})B_{x}(l_{4})\right\rangle +4\left\langle B_{x}(l_{1})B_{x}(l_{3})\right\rangle \left\langle B_{x}(l_{2})B_{x}(l_{4})\right\rangle +2\left\langle B_{x}(l_{1})B_{x}(l_{4})\right\rangle \left\langle B_{x}(l_{2})B_{x}(l_{3})\right\rangle \right)\nonumber \\
 & \rightarrow & \frac{1}{2}\frac{1}{16}g_{a\gamma}^{4}e^{-2bd}2\int_{0}^{d}dl_{1}\int_{0}^{l_{1}}dl_{2}\int_{0}^{d}dl_{3}\int_{0}^{l_{3}}dl_{4}e^{i\left(l_{1}-l_{2}\right)A+\left(l_{1}-l_{2}\right)b}e^{-i\left(l_{3}-l_{4}\right)A+\left(l_{3}-l_{4}\right)b}\left\langle B_{x}(l_{1})B_{x}(l_{2})\right\rangle \left\langle B_{x}(l_{3})B_{x}(l_{4})\right\rangle \nonumber \\
 & \rightarrow & \frac{1}{16}g_{a\gamma}^{4}\frac{\pi^{2}}{(2\pi)^{6}}\int dkk^{2}P_{B}(k)\int dk'k'{}^{2}P_{B}(k')\int_{-1}^{1}dt\left(1+t^{2}\right)\int_{-1}^{1}dt'\left(1+t'{}^{2}\right)\frac{e^{id(kt+k't')}}{\left(A-ib+kt\right)^{2}\left(A+ib-k't'\right)^{2}},\label{eq:stoB-P-ph-ph}
\end{eqnarray}
}where \(\rightarrow\) signifies that we have discarded all terms exhibiting exponential decay. The analytical expression after angular integrations over
$t$ and $t'$ in the last line is given in Eq. (\ref{eq:appen-P-sto-phph})
in Appendix C. For the sake of completeness, we remark that the ALP survival probability is dominated by the unity term, \(\mathcal{P}_{a\rightarrow a}\sim 1\), within the perturbative region.

In this work, we consider the monochromatic spectrum of the magnetic field
$P_{B}(k)=\pi^{2}B_{rms}^{2}\delta(k-k_{B})/k_{B}^{2}$, where $B_{rms}=\sqrt{\left\langle \mathbf{B}^{2}\right\rangle }$
and $\lambda_{B}=2\pi/k_{B}$ are the average strength and correlation
length of the stochastic magnetic field, respectively. We use Eqs.
(\ref{eq:stoB-P-a-ph}) and (\ref{eq:stoB-P-ph-ph}) to evaluate the
probabilities for several representative parameter sets in Fig. \ref{fig:P-stoc}.
For ALP-photon conversion probability $\mathcal{P}_{a\rightarrow\gamma}$,
when $d\ll\lambda_{\gamma}$ such that EBL attenuation effect is negligible,
there is a linear growth with propagation distance as  $\mathcal{P}_{a\rightarrow\gamma}\sim d$
for $\lambda_{B}\lesssim l_{osc}\simeq|A|^{-1}$ and the conventional
quadratic growth $\mathcal{P}_{a\leftrightarrow\gamma}\sim d^{2}$
for $\lambda_{B}\gtrsim l_{osc}$. This linear growth reflects the phenomenon of stochastic resonance in particle mixing within a stochastic background~\cite{Addazi:2024kbq,Addazi:2024mii}. In the regime \(d \gg \lambda_{\gamma}\), the conversion probability \(\mathcal{P}_{a\rightarrow\gamma}\) approaches a constant value that generally decreases with \(\lambda_{B}\), scaling approximately as \(\mathcal{P}_{a\rightarrow\gamma} \sim \lambda_{B}\).
For \(\mathcal{P}_{\gamma\rightarrow\gamma}\), we have extracted only the non-exponential-decay component (see Eq.~\eqref{eq:stoB-P-ph-ph}). Consequently, the results are reliable only at sufficiently large distances \(d \gg \lambda_{\gamma}\), where the exponential-decay contribution is strongly suppressed. This corresponds to the region above the dashed lines in Fig.~\ref{fig:P-stoc}. In the limit of large correlation length \(\lambda_B\), the photon survival probability in a stochastic magnetic field approaches that of the constant magnetic field case, as expected. As \(\lambda_B\) decreases, \(\mathcal{P}_{\gamma\rightarrow\gamma}\) is suppressed, scaling approximately as \(\mathcal{P}_{\gamma\rightarrow\gamma} \sim \lambda_B^{6}\).
Moreover, in contrast to \(\mathcal{P}_{a\rightarrow\gamma}\), which saturates at large distances, the photon survival probability decays with distance as \(\mathcal{P}_{\gamma\rightarrow\gamma} \sim 1/d^{2}\).
The different scaling behaviors of $\mathcal{P}_{a\rightarrow\gamma}$
and $\mathcal{P}_{\gamma\rightarrow\gamma}$ with respect to $\lambda_{B}$
and $d$ lead to a modified relation $\mathcal{P}_{\gamma\rightarrow\gamma}\lesssim\mathcal{P}_{a\leftrightarrow\gamma}^{2}$
in the stochastic case, in contrast to the constant magnetic field
scenario where $\mathcal{P}_{\gamma\rightarrow\gamma}\sim\mathcal{P}_{\gamma\leftrightarrow a}^{2}$
(see Figs. \ref{fig:P-const} and \ref{fig:P-stoc}). 

Motivated by the key features observed in Fig.~\ref{fig:P-stoc} and the analytical estimates of the non-exponential-decay mode derived in Appendix~B, we construct simple parametric formulas that approximate the probabilities at sufficiently large distances within the perturbative regime:
\begin{eqnarray}
\mathcal{P}_{a\rightarrow\gamma}^{\textrm{param}} & \simeq & \mathcal{O}(1)\times g_{a\gamma}^{2}B^{2}\lambda_{\gamma}^{2}\frac{1}{1+\lambda_{\gamma}^{2}A^{2}+\lambda_{\gamma}k_{B}},\nonumber \\
\mathcal{P}_{\gamma\rightarrow\gamma}^{\textrm{param}} & \simeq & \mathcal{O}(1)\times g_{a\gamma}^{4}B^{4}\lambda_{\gamma}^{4}\frac{1}{\left(1+k_{B}^{2}d^{2}\right)\left(1+\lambda_{\gamma}^{4}\left(|A|-k_{B}\right)^{4}\right)}.\label{eq:P-parameterization}
\end{eqnarray}
As as shown in Fig. \ref{fig:P-stoc}, both $\mathcal{P}_{a\rightarrow\gamma}^{\textrm{param}}$
and $\mathcal{P}_{\gamma\rightarrow\gamma}^{\textrm{param}}$ match the 
numerical results with fine accuracy. Although \(\mathcal{P}_{a\rightarrow\gamma}^{\textrm{param}}\) deviates in the resonance region \(|A| = k_B\), the discrepancy remains within an order of magnitude (\(\mathcal{O}(10)\)), which is sufficient for leading-order estimation purposes. In fact, the exact form of the non-exponential decay derived from Eq.~\eqref{eq:stoB-P-a-ph} involves an arctangent function (see Eq.~\eqref{eq:appen-P-aph-sto} in Appendix~C).

In a realistic cosmic environment, the mixing elements in Eq. (\ref{eq:EoM-const})
are given in proper units by \cite{Montanino:2017ara,Galanti:2018myb}
\begin{eqnarray}
\Delta_{\textrm{pl}} & \simeq & -10^{-11}\left(\frac{\textrm{TeV}}{\omega}\right)\left(\frac{n_{e}}{10^{-7}\textrm{cm}^{-3}}\right)\textrm{Mpc}^{-1},\nonumber \\
\Delta_{\textrm{QED}} & \simeq & 4\times10^{-8}\left(\frac{\omega}{\textrm{TeV}}\right)\left(\frac{B}{10^{-9}\textrm{Gs}}\right)^{2}\textrm{Mpc}^{-1},\nonumber \\
\Delta_{\textrm{CMB}} & \simeq & 8\times10^{-2}\left(\frac{\omega}{\textrm{TeV}}\right)\textrm{Mpc}^{-1},\nonumber \\
\Delta_{a} & \simeq & -8\left(\frac{m_{a}}{10^{-8}\textrm{eV}}\right)^{2}\left(\frac{\textrm{TeV}}{\omega}\right)\textrm{Mpc}^{-1},\nonumber \\
\Delta_{B} & \simeq & 5\times10^{-3}\left(\frac{g_{a\gamma}}{10^{-12}\textrm{GeV}^{-1}}\right)\left(\frac{B}{10^{-9}\textrm{Gs}}\right)\textrm{Mpc}^{-1}.\label{eq:mass-terms}
\end{eqnarray}
Note that $A=\Delta_{\textrm{pl}}+\Delta_{\textrm{QED}}+\Delta_{\textrm{CMB}}-\Delta_{a}$.
Here $\Delta_{\textrm{pl}}$ accounts for plasma effects in the intergalactic
medium, and $\Delta_{\textrm{QED}}$ denotes the QED vacuum polarization
correction for a photon in the magnetic background. In the intergalactic
region, both terms are typically much smaller than the ALP mass term
$\Delta_{a}$. The only contribution that can induce a sizable effect
is $\Delta_{\textrm{CMB}}$, which accounts for the electromagnetic
energy density provided by CMB \cite{Dobrynina:2014qba}. In particular,
$A$ approaches to zero at energy $\omega\simeq m_{a}/\left(10^{-9}\textrm{eV}\right)\textrm{TeV}$,
leading to a large oscillation length $l_{osc}$ as shown in left
panel in Fig. \ref{fig:PvsE}. 

Let us comment on the perturbative condition in realistic settings.
From the analysis in Appendix~B, the perturbative regime for a stochastic magnetic field requires \(\Delta_{B}^{2}\lambda_{\gamma}d \ll 1\).
This condition can be satisfied in most of the parameter space of our interests
for ALP-photon mixing over the intergalactic scale. For instance, taking
$g_{a\gamma}=10^{-12}\textrm{GeV}^{-1}$, $B_{rms}=10^{-9}\textrm{Gs}$
and $d=\textrm{Gpc}$, the perturbative condition implies $\lambda_{\gamma}\ll40\textrm{Mpc}$,
corresponding to photon energy $\omega\gtrsim10\, \textrm{TeV}$ (light
green line in left panel in Fig. \ref{fig:PvsE}). A smaller product
$g_{a\gamma}B_{rms}$ further enlarges the perturbative region. 

In Fig. \ref{fig:PvsE}, we compare the ALP-photon conversion and
photon survival probabilities in the stochastic magnetic field treated
by the simplified domain-like model (Eq. (\ref{eq:Simplied-DL}))
and the integral approach developed in this work (Eqs (\ref{eq:stoB-P-a-ph})
and (\ref{eq:stoB-P-ph-ph}) ). For comparison, the unphysical scenario of a constant magnetic field extending across the entire intergalactic region is shown as a comparative reference. For the constant magnetic field, both $\mathcal{P}_{a\rightarrow\gamma}$
and $\mathcal{P}_{\gamma\rightarrow\gamma}$ exhibit peaks caused
by the mass-equal resonance at $|A|=0$. In contrast to the simple peak structure with its well-defined resonance pattern, the scenario becomes considerably more complex when a stochastic magnetic field is introduced. 
Within the domain-like approach, both \(\mathcal{P}_{a\rightarrow\gamma}\) and \(\mathcal{P}_{\gamma\rightarrow\gamma}\) exhibit a single-peak structure similar to the constant-field case for \(m_a \gtrsim 10^{-7}\) eV. However, for \(m_a \lesssim 10^{-8}\) eV, the probabilities display a distinct feature: they decrease with increasing energy, followed by a small peak around \(\omega \simeq 100\) TeV. This feature may be attributed to the resonance condition \(|A| \simeq k_B\), i.e., \(l_{\mathrm{osc}} \simeq \lambda_B\). Such a peak in the photon survival probability, as seen in the simplified domain-like model (right panel of Fig.~\ref{fig:PvsE}), has also been observed in more sophisticated domain-like treatments~\cite{Montanino:2017ara,Galanti:2018myb}.

\begin{figure}
\includegraphics[scale=0.47]{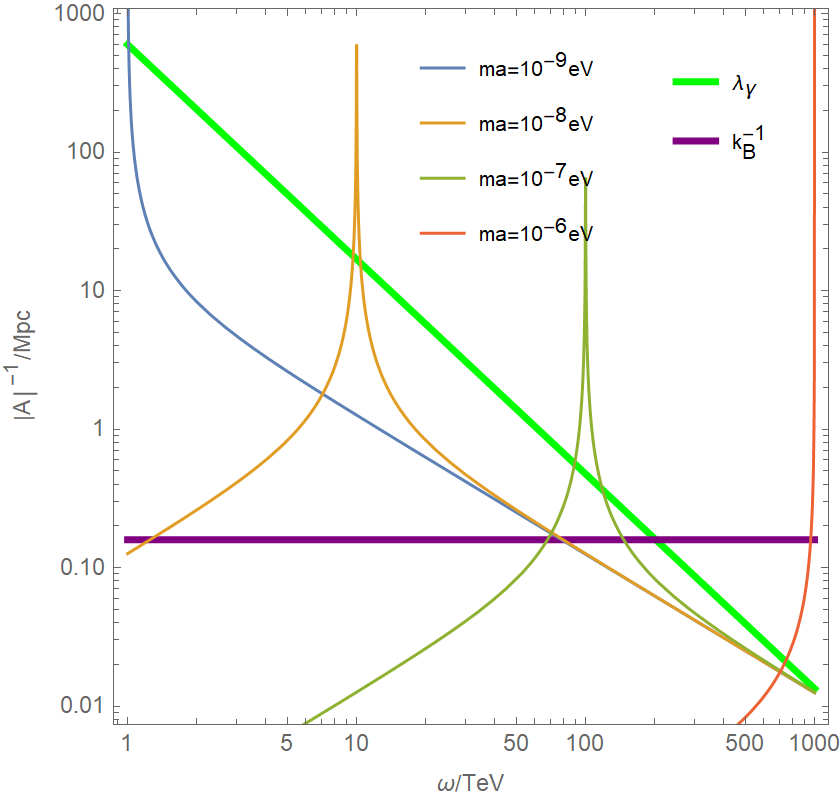}\includegraphics[scale=0.47]{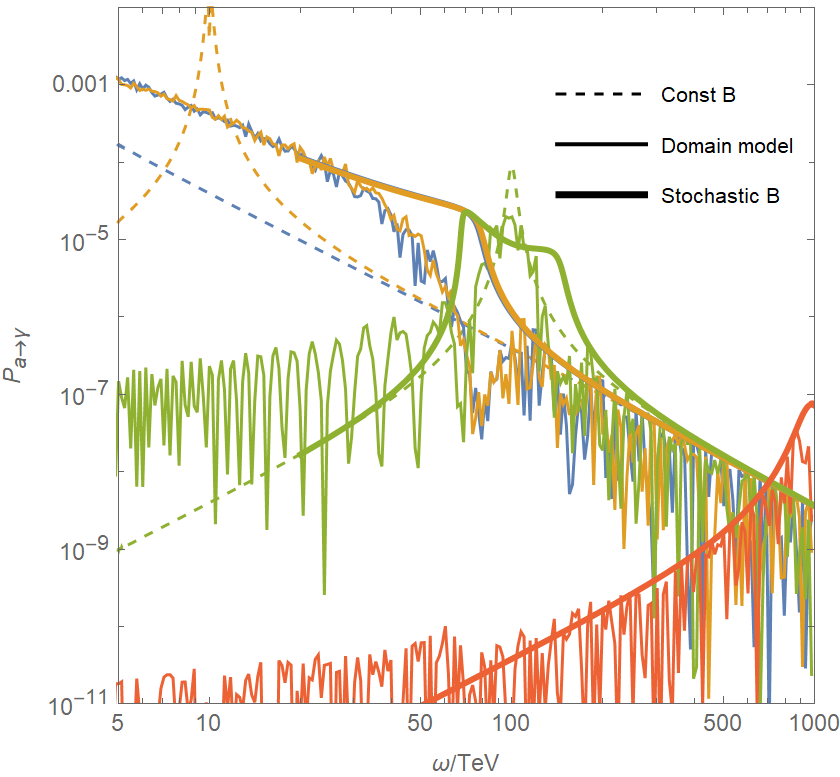}\includegraphics[scale=0.47]{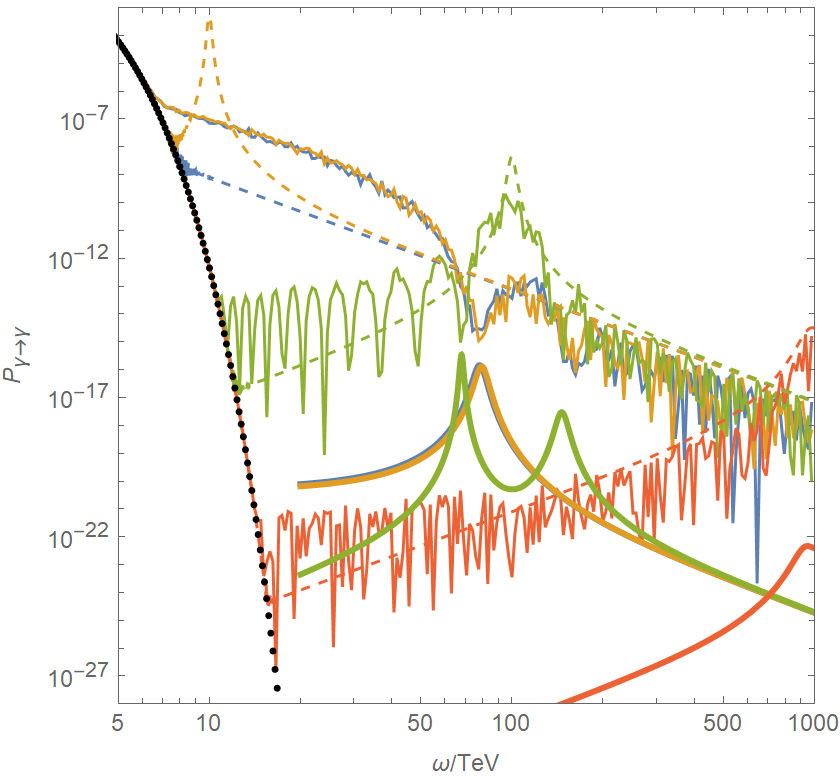}

\caption{\textbf{Left:} The characteristic length scale $1/|A|$, in the  oscillation Hamiltonian,  for different
ALP mass cases and photon mean free path $\lambda_{\gamma}$ as a
function of energy. \textbf{Middle and Right:} the ALP-photon mixing
probabilities $\mathcal{P}_{a\rightarrow\gamma}$ and $\mathcal{P}_{\gamma\rightarrow\gamma}$
in three magnetic field configurations: a constant magnetic field
filling the entire space, a simplified domain-like model, and a stochastic
magnetic field. The parameters are set to $B_{rms}=10^{-9}\textrm{Gs}$,
$g_{a\gamma}=10^{-12}\textrm{GeV}^{-1}$ and $d=\textrm{Gpc}$. For
the stochastic magnetic field, its distribution is Gaussian and the
spectrum is monochromatic with correlation length $\lambda_{B}=1\textrm{Mpc}$.
The black dotted line in right panel denotes the photon survival probability
in the absence of ALP mixing. The thick curves are constrained to
$\omega\gtrsim20\textrm{TeV}$ where the perturbative condition is well satisfied in the stochastic case.}

\label{fig:PvsE}
\end{figure}

For the integral approach, we again assume a monochromatic spectrum for the Gaussian magnetic field and present the results (thick curves) within the perturbative regime, as shown in Fig.~\ref{fig:PvsE}. The amplitude of $\mathcal{P}_{a\rightarrow\gamma}$
is comparable to that obtained in domain-like model across the considered
mass cases. Combining the left panel in Fig. \ref{fig:PvsE} with
upper row in Fig. \ref{fig:P-stoc}, the peak or knee structure in
$\mathcal{P}_{a\rightarrow\gamma}$ originates from the stochastic
resonance when $|A|^{-1}\gtrsim k_{B}^{-1}$, i.e. $\lambda_{B}\lesssim l_{osc}$.
For photon survival probability, two notable features emerge: i) $\mathcal{P}_{\gamma\rightarrow\gamma}$ is suppressed by
several orders of magnitude relative to the domain-like model; ii) the peak structures of all considered masses cases could
be explained by the resonance at $|A|\simeq k_{B}$. In particular,
a clear double-peak structure emerges  for $m_{a}=10^{-7}$eV. 

As shown in Fig.~\ref{fig:PvsE}, within the perturbative regime the domain-like treatment yields the same quadratic amplitude relation as in the constant-field case, namely \(\mathcal{P}_{\gamma\rightarrow\gamma} \simeq \mathcal{P}_{a\rightarrow\gamma}^{2}\). This behavior can be interpreted as arising from a double oscillation process \(\gamma \rightarrow a \rightarrow \gamma\). In contrast, the integral approach reveals that this quadratic relation no longer holds, and the peak structures of \(\mathcal{P}_{a\rightarrow\gamma}\) and \(\mathcal{P}_{\gamma\rightarrow\gamma}\) differ significantly.
By consistently preserving the stochastic nature of the magnetic field from the outset, the integral approach captures the nontrivial interplay between EBL attenuation and ALP-photon mixing in a stochastic magnetic environment. These results highlight the importance of moving beyond simplified domain-like approximations when modeling magnetic turbulence and motivate further investigation in future work. 

\section{Non-Gaussian magnetic field}

\label{sec:Non-Gaussian-magnetic-field}

In the previous section, we considered a Gaussian random magnetic field. From now on, we consider the case of a non-Gaussian distribution. In this case, the four-point correlation function of the magnetic field can no longer be fully decomposed into products of two-point functions via Wick's theorem (Eq.~\eqref{eq:4PCF-Wick}). Instead, it includes an additional nonvanishing connected component, \(\left\langle B(l_{1})B(l_{2})B(l_{3})B(l_{4})\right\rangle _{c}\), which corresponds to the trispectrum in momentum space. A detailed treatment of the trispectrum is challenging, as its specific form depends on the underlying magnetogenesis mechanism—whether of primordial origin or generated by astrophysical processes. To qualitatively assess the impact of magnetic non-Gaussianity on the photon survival probability in ALP-photon mixing, we therefore adopt three heuristic approaches. 

For a qualitative analysis, we treat the magnetic vector field simply as a scalar in three-dimensional space, denoted \(B(\boldsymbol{x})\). Following the derivation leading to Eqs.~\eqref{eq:stoB-P-ph-ph}, the non-exponential-decay contribution arises from an integral of the form \(\int_{0}^{d}dl_{1}\int_{0}^{l_{1}}dl_{2}\int_{0}^{d}dl_{3}\int_{0}^{l_{3}}dl_{4}\). From this, the relevant part of \(\mathcal{P}_{\gamma\rightarrow\gamma}\) can be extracted as {\small
\begin{eqnarray}
\mathcal{P}_{\gamma\rightarrow\gamma}^{\textrm{NG}} & = & 4\frac{1}{2}e^{-2bd}\frac{1}{16}g_{a\gamma}^{4}\int_{0}^{d}dl_{1}\int_{0}^{l_{1}}dl_{2}\int_{0}^{d}dl_{3}\int_{0}^{l_{3}}dl_{4}e^{i\left(l_{1}-l_{2}\right)A+\left(l_{1}-l_{2}\right)b}e^{-i\left(l_{3}-l_{4}\right)A+\left(l_{3}-l_{4}\right)b}\left\langle B(l_{1})B(l_{2})B(l_{3})B(l_{4})\right\rangle _{c},\label{eq:NG-P-general}
\end{eqnarray}
}where the overall prefactor $4$ accounts for scalar approximation
$B_{x}(\boldsymbol{x})=B_{y}(\boldsymbol{x})=B(\boldsymbol{x})$.

\begin{figure}
\includegraphics[scale=0.47]{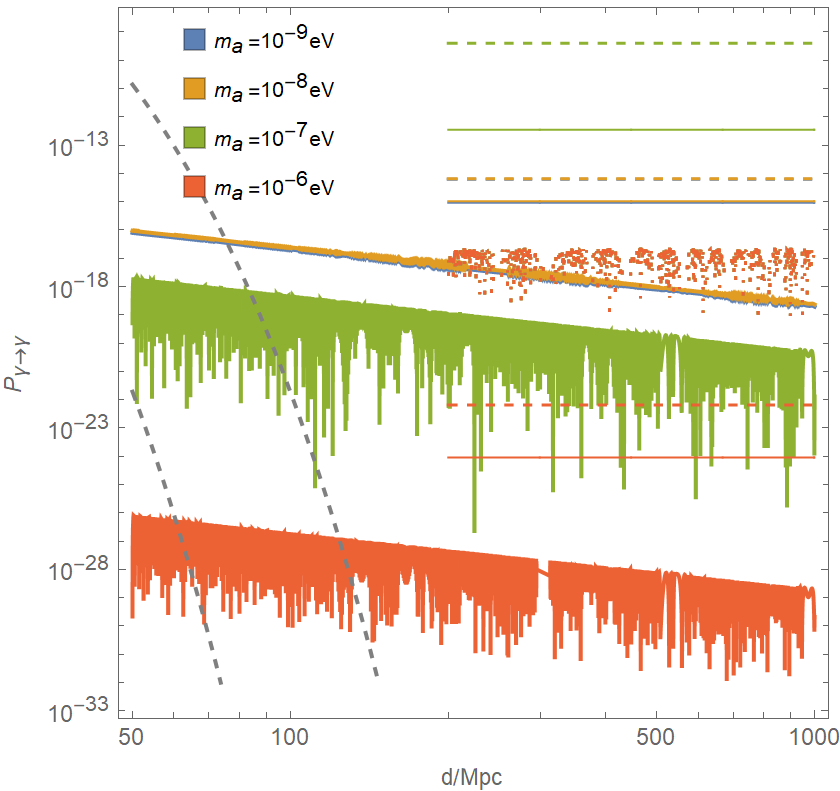}\includegraphics[scale=0.47]{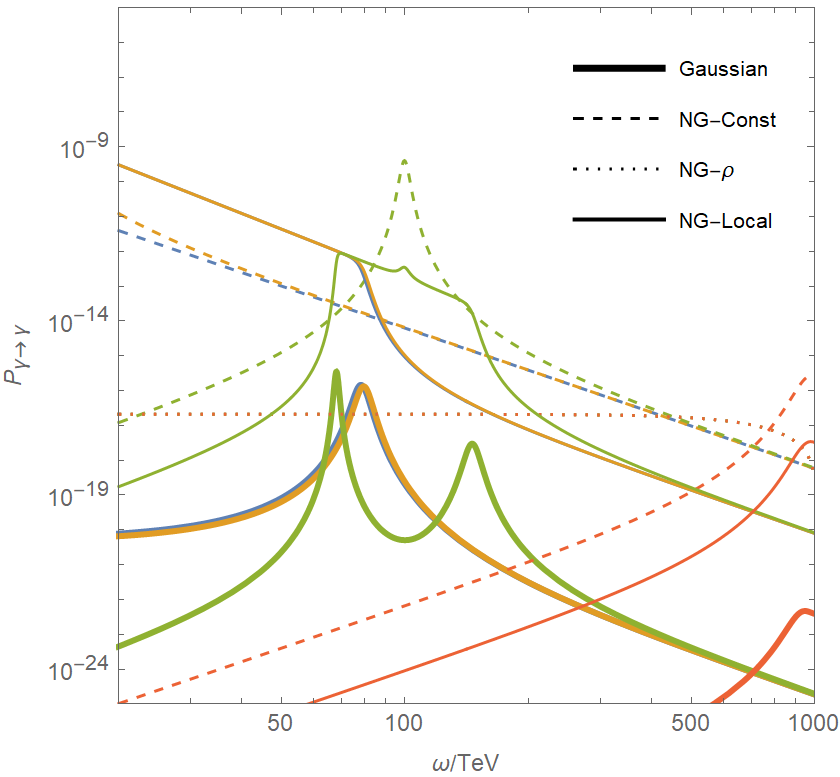}\includegraphics[scale=0.47]{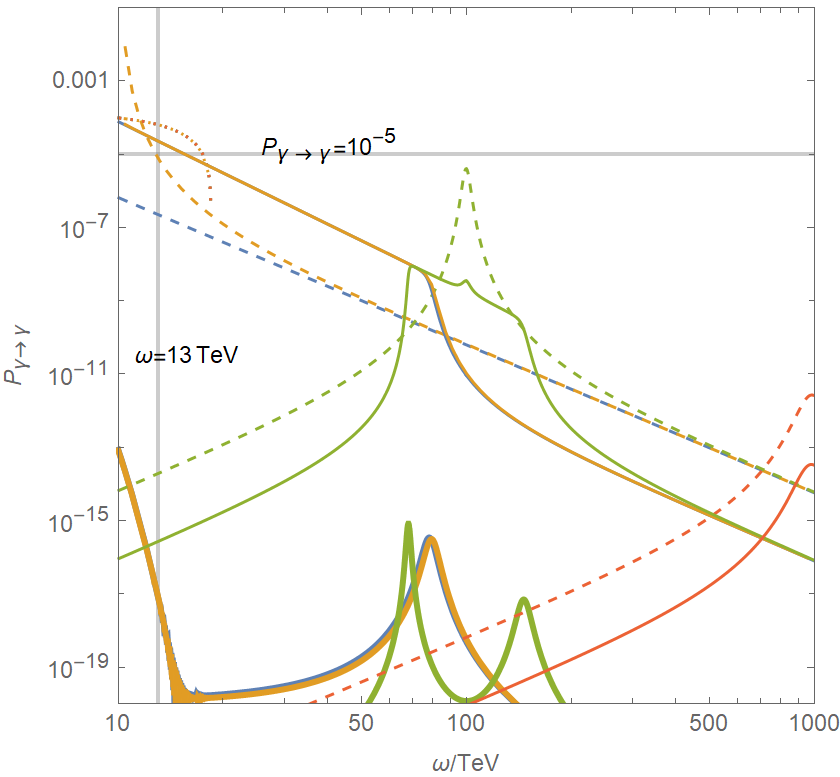} 

\caption{The photon survival probability $\mathcal{P}_{\gamma\rightarrow\gamma}$ in the stochastic magnetic field with a monochromatic spectrum with $\lambda_{B}=1\, \textrm{Mpc}$, for
Gaussian (thick line) and non-Gaussian distributions (case-1 for dashed
line, case-2 for dotted line and case-3 for thin line). The left panel
is at fixed $\omega=100\textrm{TeV}$, and the middle and right  panels is fixed at $d=\textrm{Gpc}$. The middle panel corresponds to the weak non-Gaussianity, where parameters are set to $\kappa=0.01$ for case-1, $\kappa=0.01,\lambda_{\rho}=0.1\textrm{Mpc}$  for case-2, and $g_{NL}=0.01$ for case-3. The right panel corresponds  to strong non-Gaussianity, where parameters are set to $\kappa=100$ for case-1, $\kappa=1, \lambda_{\rho}=50\textrm{Mpc}$ for case-2, and $g_{NL}=100$ for case-3. Note that non-gaussianities  highly boost the photon survival probability with model dependent profiles. }

\label{fig:NGaussianB}
\end{figure}

\vspace{0.2cm}

\textbf{\textit{Non-Gaussian case 1}}. The simplest case is to assume a constant 4-point function 
\begin{equation}
    \left\langle B(l_{1})B(l_{2})B(l_{3})B(l_{4})\right\rangle _{c}=\kappa B_{rms}^{4}\, . 
\end{equation}
Under this assumption, we obtain 
\begin{eqnarray}
\mathcal{P}_{\gamma\rightarrow\gamma}^{\textrm{NG}} & = & 4\frac{1}{2}e^{-2db}\frac{1}{16}g_{a\gamma}^{4}\kappa B_{rms}^{4}\int_{0}^{d}dl_{1}\int_{0}^{l_{1}}dl_{2}\int_{0}^{d}dl_{3}\int_{0}^{l_{3}}dl_{4}e^{i\left(l_{1}-l_{2}\right)A+\left(l_{1}-l_{2}\right)b}e^{-i\left(l_{3}-l_{4}\right)A+\left(l_{3}-l_{4}\right)b}\nonumber \\
 & \rightarrow & \frac{1}{8}\kappa g_{a\gamma}^{4}B_{rms}^{4}\frac{1}{(A^{2}+b^{2})^{2}},
\end{eqnarray}
where $\rightarrow$ denotes the non exponential-decay part $(db<  1)$. The parameter
$\kappa$ is the kurtosis of magnetic field, $\kappa=\left\langle B^{4}\right\rangle /\left\langle B^{2}\right\rangle ^{2}-3$,
which is commonly used to quantify intermittency in MHD simulations
\cite{Kowal:2006jw}.

\vspace{0.2cm}

\textbf{\textit{Non-Gaussian case 2}}. In the literature, magnetic non-Gaussianity is often characterized via the magnetic energy density \(\rho(\boldsymbol{x}) = B^{2}(\boldsymbol{x})\) or the stress-energy tensor~\cite{Brown:2005kr,Sai:2025cry}. Adopting a similar strategy, we introduce an ansatz that incorporates a Dirac delta function by hand:
\begin{eqnarray}
\left\langle B(l_{1})B(l_{2})B(l_{3})B(l_{4})\right\rangle _{c} & = & \left\langle \rho(l_{1})\rho(l_{2})\right\rangle _{c}\delta(l_{1}/l_{3}-1)\delta(l_{2}/l_{4}-1)\nonumber  +\left\langle \rho(l_{1})\rho(l_{3})\right\rangle _{c}\delta(l_{1}/l_{2}-1)\delta(l_{3}/l_{4}-1)\nonumber \\
 &  & +\left\langle \rho(l_{1})\rho(l_{2})\right\rangle _{c}\delta(l_{1}/l_{4}-1)\delta(l_{2}/l_{3}-1),\label{eq:NGcase2-B}
\end{eqnarray}
where $\left\langle \rho(l_{1})\rho(l_{2})\right\rangle _{c}=\left\langle \rho(l_{1})\rho(l_{2})\right\rangle -\left\langle \rho(l_{1})\right\rangle \left\langle \rho(l_{2})\right\rangle $
is the connected two point correlation function of the magnetic energy
density. This quantity characterizes the spatial distribution of magnetic energy. A long-range connected correlation \(\left\langle \rho\rho\right\rangle _{c}\) indicates a relatively smooth and homogeneous field configuration, whereas a short-range correlation corresponds to a tangled and intermittent field. In general, fluctuations in the magnetic energy decay rapidly beyond the intrinsic correlation length of the magnetic field. In momentum space, the power spectrum of the magnetic energy is defined as
\begin{eqnarray}
\left\langle \rho(\boldsymbol{x}_{1})\rho(\boldsymbol{x}_{2})\right\rangle _{c} & = & \int d^{3}\boldsymbol{k}e^{i\left(\boldsymbol{x}_{1}-\boldsymbol{x}_{2}\right)\cdot\boldsymbol{k}}P_{\rho}(\boldsymbol{k}).\label{eq:rhoCF}
\end{eqnarray}
Substituting Eqs. (\ref{eq:NGcase2-B}) and (\ref{eq:rhoCF}) into
Eq. (\ref{eq:NG-P-general}), we find that only the combination $\delta(l_{1}/l_{3}-1)\delta(l_{2}/l_{4}-1)$
contributes to the non-exponential-decay part
\begin{eqnarray}
\mathcal{P}_{\gamma\rightarrow\gamma}^{\textrm{NG}} & \rightarrow & 4\frac{1}{2}e^{-2db}\frac{1}{16}g_{a\gamma}^{4}\int_{0}^{d}dl_{1}\int_{0}^{l_{1}}dl_{2}\int_{0}^{d}dl_{3}\int_{0}^{l_{3}}dl_{4}e^{2\left(l_{1}-l_{2}\right)b}\left\langle \rho(l_{1})\rho(l_{2})\right\rangle _{c}\delta(l_{1}/l_{3}-1)\delta(l_{2}/l_{4}-1)\nonumber \\
 & \rightarrow & \frac{\pi}{4}g_{a\gamma}^{4}\int dkk^{2}P_{\rho}(k)\int_{-1}^{1}dt\frac{-e^{idkt}(1-2bd-idkt)}{(2b+ikt)^{4}}.\label{eq:NGcase2-P}
\end{eqnarray}
Its expression after integration over $t$ is given in Eq. (\ref{eq:appen-NG2})
in Appendix C. In this work, we consider a monochromatic spectrum
of the magnetic energy density, $P_{\rho}(k)=\kappa B_{rms}^{4}\delta(k-k_{\rho})/k^{2}$
with characteristic wavelength $\lambda_{\rho}=2\pi/k_{\rho}$. We
set $\lambda_{\rho}=0.1$Mpc for intergalactic magnetic field with
typical $\lambda_{B}=1$Mpc. 

\vspace{0.1cm}

\textbf{\textit{Non-Gaussian case 3}}. We follow the standard perturbative approach employed in the analysis of non-Gaussianity in primordial curvature perturbations and the CMB trispectrum. To this end, we first rescale \(B\) to a dimensionless quantity \(\mathcal{B}(\boldsymbol{x}) = B(\boldsymbol{x}) / B_{\mathrm{rms}}\). A common parametrization of non-Gaussianity is given by a local expansion in terms of underlying Gaussian random fields~\cite{Seery:2008ax,ade2014planck}:
\begin{eqnarray}
\mathcal{B}(\boldsymbol{x}_{1}) & = & \mathcal{B}_{G}+\frac{1}{2}\sqrt{\tau_{NL}}\left(\mathcal{B}_{G}^{2}-\left\langle \mathcal{B}_{G}^{2}\right\rangle \right)+\frac{9}{25}g_{NL}\mathcal{B}_{G}^{3}+...,
\end{eqnarray}
where $\tau_{NL}$ and $g_{NL}$ are dimensionless constants. The
connected 4-point correlation function is
\begin{eqnarray}
\left\langle \mathcal{B}(\boldsymbol{x}_{1})\mathcal{B}(\boldsymbol{x}_{2})\mathcal{B}(\boldsymbol{x}_{3})\mathcal{B}(\boldsymbol{x}_{4})\right\rangle _{c} & = & \frac{1}{(2\pi)^{12}}\int d^{3}\boldsymbol{k}_{1}\int d^{3}\boldsymbol{k}_{2}\int d^{3}\boldsymbol{k}_{3}\int d^{3}\boldsymbol{k}_{4}(2\pi)^{3}\delta^{(3)}\left(\boldsymbol{k}_{1}+\boldsymbol{k}_{2}+\boldsymbol{k}_{3}+\boldsymbol{k}_{4}\right)\nonumber \\
 &  & \times e^{i\left(\boldsymbol{k}_{1}\cdot\boldsymbol{x}_{1}+\boldsymbol{k}_{2}\cdot\boldsymbol{x}_{2}+\boldsymbol{k}_{3}\cdot\boldsymbol{x}_{3}+\boldsymbol{k}_{4}\cdot\boldsymbol{x}_{4}\right)}T(\boldsymbol{k}_{1},\boldsymbol{k}_{2},\boldsymbol{k}_{3},\boldsymbol{k}_{4})
\end{eqnarray}
with 
\begin{eqnarray}
T(\boldsymbol{k}_{1},\boldsymbol{k}_{2},\boldsymbol{k}_{3},\boldsymbol{k}_{4}) & = & \tau_{NL}\underset{a\neq b,c\&b<c}{\sum}P_{\mathcal{B}}(k_{ab})P_{\mathcal{B}}(k_{b})P_{\mathcal{B}}(k_{c})+\frac{54}{25}g_{NL}\underset{a<b<c}{\sum}P_{\mathcal{B}}(k_{a})P_{\mathcal{B}}(k_{b})P_{\mathcal{B}}(k_{c}).
\end{eqnarray}
We consider the simplest case: the collinear configuration, in which all wave vectors have equal magnitude and are aligned either parallel or antiparallel to one another~\cite{Trivedi:2013wqa}. This choice implicitly assumes that the collinear configuration provides the dominant contribution to the trispectrum in a certain class of magnetogenesis models. To describe this configuration, we decompose the Dirac delta function as
\begin{eqnarray}
(2\pi)^{3}\delta^{(3)}\left(\boldsymbol{k}_{1}+\boldsymbol{k}_{2}+\boldsymbol{k}_{3}+\boldsymbol{k}_{4}\right)_{\textrm{collinear}} & = & \delta^{(2)}\left(\boldsymbol{\varOmega}_{2}-\boldsymbol{\varOmega}_{1}\right)\delta^{(2)}\left(\boldsymbol{\varOmega}_{3}+\boldsymbol{\varOmega}_{1}\right)(2\pi)^{3}\delta^{(3)}\left(\boldsymbol{k}_{1}+\boldsymbol{k}_{2}+\boldsymbol{k}_{3}+\boldsymbol{k}_{4}\right)\nonumber \\
 &  & +\delta^{(2)}\left(\boldsymbol{\varOmega}_{2}+\boldsymbol{\varOmega}_{1}\right)\delta^{(2)}\left(\boldsymbol{\varOmega}_{3}-\boldsymbol{\varOmega}_{1}\right)(2\pi)^{3}\delta^{(3)}\left(\boldsymbol{k}_{1}+\boldsymbol{k}_{2}+\boldsymbol{k}_{3}+\boldsymbol{k}_{4}\right)\nonumber \\
 &  & +\delta^{(2)}\left(\boldsymbol{\varOmega}_{2}+\boldsymbol{\varOmega}_{1}\right)\delta^{(2)}\left(\boldsymbol{\varOmega}_{3}+\boldsymbol{\varOmega}_{1}\right)(2\pi)^{3}\delta^{(3)}\left(\boldsymbol{k}_{1}+\boldsymbol{k}_{2}+\boldsymbol{k}_{3}+\boldsymbol{k}_{4}\right),
\end{eqnarray}
where $\boldsymbol{\varOmega}$ denotes the $\left(\theta,\varphi\right)$-direction
of the momentum vector in the unit 2-sphere, and $\delta^{(2)}\boldsymbol{\varOmega}=\delta(\theta)\delta(\varphi)/\textrm{sin}\theta$. 

Again, we adopt a monochromatic spectrum for the rescaled magnetic field, given by \(P_{\mathcal{B}}(\boldsymbol{k}) = P_{B}(k)/B_{\mathrm{rms}}^{2} = \pi^{2}\delta(k - k_{B})/k_{B}^{2}\). In the collinear configuration, only the \(g_{NL}\) term in the trispectrum \(T(\boldsymbol{k}_{1},\boldsymbol{k}_{2},\boldsymbol{k}_{3},\boldsymbol{k}_{4})\) contributes. Combining these elements, the connected four-point correlation function in the collinear configuration takes the form {\small
\begin{eqnarray}
\left\langle \mathcal{B}(l_{1})\mathcal{B}(l_{2})\mathcal{B}(l_{3})\mathcal{B}(l_{4})\right\rangle _{c} & = & \frac{1}{(2\pi)^{12}}\int d^{3}\boldsymbol{k}_{1}\int d^{3}\boldsymbol{k}_{2}\int d^{3}\boldsymbol{k}_{3}\int d^{3}\boldsymbol{k}_{4}(2\pi)^{3}\delta^{(3)}\left(\boldsymbol{k}_{1}+\boldsymbol{k}_{2}+\boldsymbol{k}_{3}+\boldsymbol{k}_{4}\right)\nonumber \\
 &  & \times T(\boldsymbol{k}_{1},\boldsymbol{k}_{2},\boldsymbol{k}_{3},\boldsymbol{k}_{4})e^{i\left(k_{1}l_{1}\textrm{cos}\theta_{1}+k_{2}l_{2}\textrm{cos}\theta_{2}+k_{3}l_{3}\textrm{cos}\theta_{3}+k_{4}l_{4}\textrm{cos}\theta_{4}\right)}\nonumber \\
 & = & \frac{1}{(2\pi)^{9}}\int k_{1}^{2}dk_{1}\int d\boldsymbol{\varOmega}_{1}\int k_{2}^{2}dk_{2}\int d\boldsymbol{\varOmega}_{2}\int k_{3}^{2}dk_{3}\int d\boldsymbol{\varOmega}_{3}T(\boldsymbol{k}_{1},\boldsymbol{k}_{2},\boldsymbol{k}_{3},\boldsymbol{k}_{4}=-\boldsymbol{k}_{1}-\boldsymbol{k}_{2}-\boldsymbol{k}_{3})\nonumber \\
 &  & e^{i\left(k_{1}\left(l_{1}-l_{4}\right)\textrm{cos}\theta_{1}+k_{2}(l_{2}-l_{4})\textrm{cos}\theta_{2}+k_{3}(l_{3}-l_{4})\textrm{cos}\theta_{3}\right)}\left\{ \delta^{(2)}\left(\boldsymbol{\varOmega}_{2}-\boldsymbol{\varOmega}_{1}\right)\delta^{(2)}\left(\boldsymbol{\varOmega}_{3}+\boldsymbol{\varOmega}_{1}\right)\right.\nonumber \\
 &  & \left.+\delta^{(2)}\left(\boldsymbol{\varOmega}_{2}+\boldsymbol{\varOmega}_{1}\right)\delta^{(2)}\left(\boldsymbol{\varOmega}_{3}-\boldsymbol{\varOmega}_{1}\right)+\delta^{(2)}\left(\boldsymbol{\varOmega}_{2}+\boldsymbol{\varOmega}_{1}\right)\delta^{(2)}\left(\boldsymbol{\varOmega}_{3}+\boldsymbol{\varOmega}_{1}\right)\right\} \nonumber \\
 & = & \frac{1}{2^{6}\pi^{2}}\frac{54}{25}g_{NL}\int_{-1}^{1}dt\left\{ e^{ik_{B}t\left(l_{1}+l_{2}-l_{3}-l_{4}\right)}+e^{ik_{B}t\left(l_{1}-l_{2}+l_{3}-l_{4}\right)}+e^{ik_{B}t\left(l_{1}-l_{2}-l_{3}+l_{4}\right)}\right\} .
\end{eqnarray}
}Plugging it into Eq. (\ref{eq:NG-P-general}) yields {\small
\begin{eqnarray}
\mathcal{P}_{\gamma\rightarrow\gamma}^{\textrm{NG}} & = & 4\frac{1}{2}e^{-2bd}\frac{1}{16}g_{a\gamma}^{4}B_{rms}^{4}\int_{0}^{d}dl_{1}\int_{0}^{l_{1}}dl_{2}\int_{0}^{d}dl_{3}\int_{0}^{l_{3}}dl_{4}e^{i\left(l_{1}-l_{2}\right)A+\left(l_{1}-l_{2}\right)b}e^{-i\left(l_{3}-l_{4}\right)A+\left(l_{3}-l_{4}\right)b}\left\langle \mathcal{B}(l_{1})\mathcal{B}(l_{2})\mathcal{B}(l_{3})\mathcal{B}(l_{4})\right\rangle _{c}\nonumber \\
 & \rightarrow & \frac{1}{2^{9}\pi^{2}}g_{a\gamma}^{4}B_{rms}^{4}\frac{54}{25}g_{NL}\int_{-1}^{+1}dt\left(\frac{1}{A^{4}+2A^{2}(b^{2}-k_{B}^{2}t^{2})+(b^{2}+k_{B}^{2}t^{2})^{2}}+\frac{e^{2ik_{B}td}}{(A^{2}+(b+ik_{B}t)^{2})^{2}}+\frac{1}{(b^{2}+(A+k_{B}t)^{2})^{2}}\right),\label{eq:NGcase3-P}
\end{eqnarray}
} whose expression after integration over $t$ is given in Eq. (\ref{eq:Appen-NGcase3})
in Appendix C. 

We mention that in above three non-Gaussian cases, $\mathcal{P}_{\gamma\rightarrow\gamma}^{\textrm{NG}}$
is guaranteed to be real, though not necessarily positive definite.
In the parameter space of interest, we find that cases 1 and 3 remain
positive, whereas case 2 can be either positive or negative.

In Fig.~\ref{fig:NGaussianB}, we compare the photon survival probability in stochastic magnetic fields with Gaussian and non-Gaussian distributions. The left panel shows that the three simple non-Gaussian models considered above yield a distance-independent \(\mathcal{P}_{\gamma\rightarrow\gamma}\). The right panel demonstrates that in non-Gaussian Cases 1 and 3, \(\mathcal{P}_{\gamma\rightarrow\gamma}\) exceeds the corresponding Gaussian result by several orders of magnitude across the entire energy range for each fixed \(m_a\). A distinctive feature of non-Gaussian Case 2 is that \(\mathcal{P}_{\gamma\rightarrow\gamma}\) becomes nearly independent of \(m_a\), exhibiting a flat spectrum. This leads to a broad enhancement over the Gaussian result, in contrast to the mass-dependent behavior observed in Cases 1 and 3. Furthermore, the right panel shows that in a strongly non-Gaussian regime, the photon survival probability can reach \(10^{-4}\) at 13\,TeV, potentially accounting for the unexpected photon event observed by LHAASO.

\section{Discussions and Remarks}

\label{sec:Discussions and Remarks}

The mixing of ALPs and photons in the presence of the EBL was analyzed, considering various magnetic field models: constant, Gaussian stochastic, and non-Gaussian. 
For constant fields, we derived exact analytical expressions for all conversion probabilities and identified the perturbative regime corresponding to weak magnetic fields ($g_{a\gamma}B\lambda_\gamma \ll 1$) in the strong attenuation region $d \gg \lambda_\gamma$. 
Within this regime, the ALP-photon conversion probability scales as $\mathcal{P}_{a\rightarrow\gamma} \sim B^2$, while the photon survival probability scales as $\mathcal{P}_{\gamma\rightarrow\gamma} \sim B^4$.

For Gaussian stochastic fields, we developed an integral perturbation framework and obtained analytical expressions for the non-exponential-decay contributions in the strong-attenuation regime. 
We also provided simple parametrized formulas adequate for order-of-magnitude estimates. 
To compare with the widely used domain-like model, we found that while ALP-photon conversion probabilities remain comparable, the photon survival probability is suppressed by several orders of magnitude. 
Both probabilities exhibit distinct multi-peak structures arising from the combined effects of mass-equal resonance ($\Delta_a = \Delta_{\textrm{CMB}}$), stochastic resonance ($\lambda_B \lesssim l_{\mathrm{osc}}$), and EBL attenuation. 
Extending to non-Gaussian magnetic fields via three ad-hoc models (constant kurtosis, energy-density correlation, and local trispectrum expansion), we demonstrated that non-Gaussianity can enhance the photon survival probability by orders of magnitude relative to the Gaussian case. 
A more rigorous treatment of particle mixing in non-Gaussian magnetic fields is left for future work.

\begin{figure}
\includegraphics[scale=0.46]{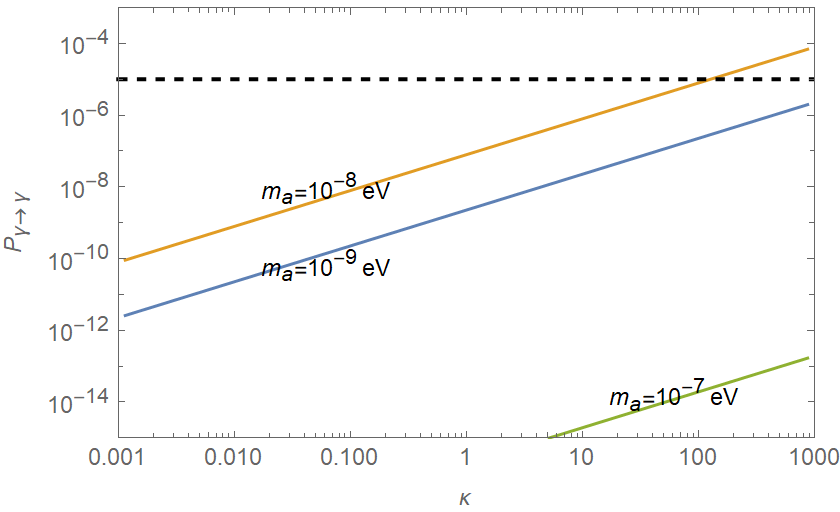}\includegraphics[scale=0.46]{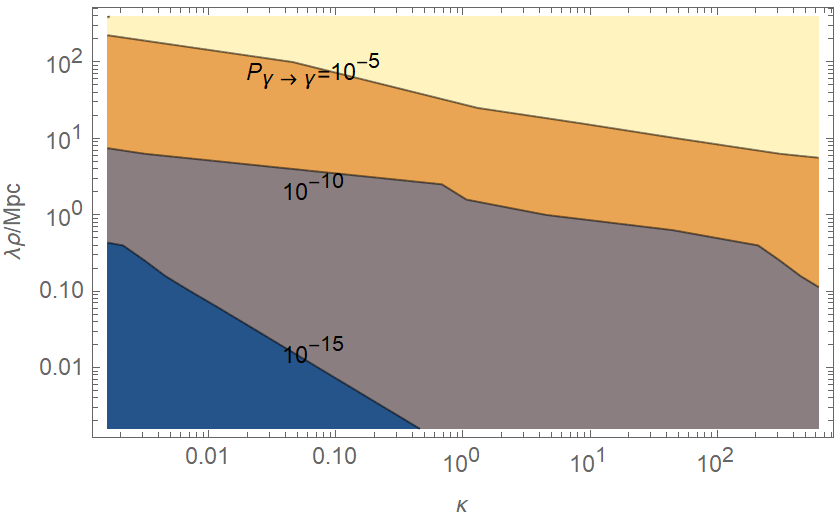}\includegraphics[scale=0.46]{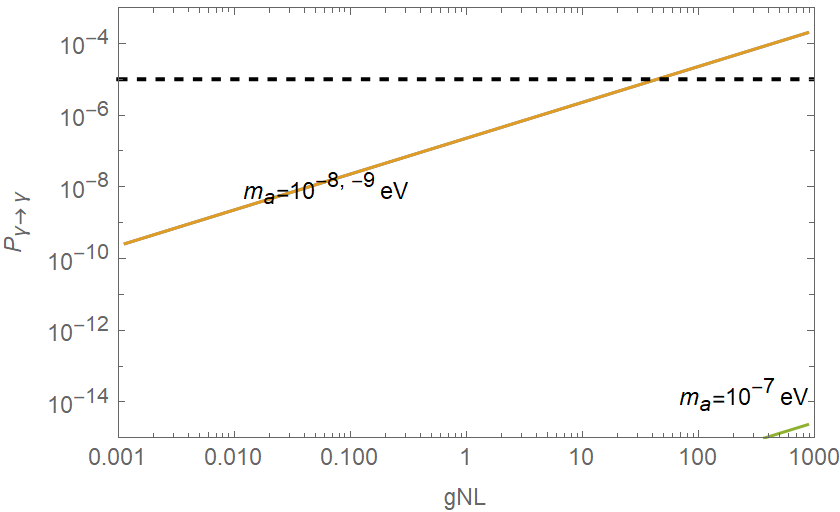}
\caption{The photon survival probability $\mathcal{P}_{\gamma\rightarrow\gamma}$ in the non-Gaussian magnetic field as a function of corresponding non-Gaussian parameters in Case 1 (left), Case 2 (right) and Case 3 (right). In Case 2 the probability depends on two non-Gaussian parameters $\kappa$ and $\lambda_\rho$, whereas independent of ALP mass. The dashed black line $\mathcal{P}_{\gamma\rightarrow\gamma}=10^{-5}$ denotes the marginal required probability to explain the 13 TeV photon observed by LHAASO. Such a high energetic photon can be explained by ALP-photon mixing in a strong non-Gaussian magnetic background. }

\label{fig:NG123}
\end{figure}

We now comment on a more realistic astrophysical scenario. Gamma rays emitted from an extragalactic source propagate through intergalactic space before reaching the Milky Way and Earth. In the presence of ALP-photon oscillations, the observed and emitted photon fluxes are related by
\begin{equation}
F_{\gamma}^{\textrm{earth}} =F_{\gamma}^{\textrm{source}} \mathcal{P}_{\gamma\rightarrow\gamma}^{\textrm{tot}}  \simeq  F_{\gamma}^{\textrm{source}}\left(\mathcal{P}_{\gamma\rightarrow\gamma}^{\textrm{IG}} + \mathcal{P}_{a\leftrightarrow\gamma}^{\textrm{IG}}\mathcal{P}_{a\leftrightarrow\gamma}^{\textrm{Gal}} + \left(\mathcal{P}_{a\leftrightarrow\gamma}^{\textrm{Gal}}\right)^2\right),\label{eq:Fearth-disc}
\end{equation}
where "IG" denotes the intergalactic region and "Gal" denotes both the source host galaxy and the Milky Way, assuming similar small conversion probabilities $\mathcal{P}_{a\leftrightarrow\gamma}^{\textrm{Host}} \simeq \mathcal{P}_{a\leftrightarrow\gamma}^{\textrm{MW}} \simeq \mathcal{P}_{a\leftrightarrow\gamma}^{\textrm{Gal}} \ll 1$. For a typical coupling $g_{a\gamma} \simeq 10^{-12}\,\textrm{GeV}^{-1}$, EBL attenuation in galaxies is negligible below the PeV scale, yielding $\mathcal{P}_{a\leftrightarrow\gamma}^{\textrm{Gal}} \simeq 10^{-4}$ for $B \simeq 10^{-6}\,\textrm{Gs}$ and $d \simeq 10\,\textrm{kpc}$. In intergalactic space, for $B \simeq 10^{-9}\,\textrm{Gs}$ and $d \simeq \textrm{Gpc}$, our results show $\mathcal{P}_{a\leftrightarrow\gamma}^{\textrm{IG}} \lesssim 10^{-4}$ and $\mathcal{P}_{\gamma\rightarrow\gamma}^{\textrm{IG}} \lesssim 10^{-15}$, potentially enhanced to $\mathcal{P}_{\gamma\rightarrow\gamma}^{\textrm{IG}} \simeq 10^{-10}$ with weak non-Gaussianity (see Fig. \ref{fig:NGaussianB}). Therefore, for typical parameters in realistic astrophysical environment, the dominant contribution to the observed photon flux arises from ALP-photon conversion inside the host galaxy and the Milky Way. To highlight the impact of EBL attenuation on ALP mixing, three avenues can be explored: (i) energies beyond the PeV scale, where EBL effects inside galaxies become significant; (ii) sources not embedded in galactic environments, such as primordial black hole evaporation or Exotic PeVatrons emissions \cite{Addazi:2025ldl} before galaxy formation; and (iii)  in the strong non-Gaussian regime. 

We comment on point (iii) in particular. To explain the unexpected high energetic photon with energy larger than $10$TeV observed by LHAASO, the total photon survival probability from source to earth $\mathcal{P}_{\gamma\rightarrow\gamma}^{\textrm {tot}}$ is required to be marginally   larger than  $10^{-5}$  \cite{LHAASA:2023pay, Galanti:2022chk}. Generally speaking, $\mathcal{P}_{\gamma\rightarrow\gamma}^{\textrm {tot}}$ is dominated by the ALP-photon oscillation inside the galaxies due to their relatively stronger magnetic field ($\sim 10^{-5}$Gs) compared to the intergalactic one ($\sim 10^{-9}$Gs). However, from Fig \ref{fig:NGaussianB}, we observe that non-Gaussianity of magnetic field can boost the photon survival probability in the intergalactic region. We scan a broad non-Gaussian parameter space in three cases analyzed in last section and obtain Fig. \ref{fig:NG123}. It demonstrates that photon survival probability in the intergalactic region $\mathcal{P}_{\gamma\rightarrow\gamma}^{\textrm {IG}}$ can reach to  $ \sim 10^{-5}$ in very strong non-Gaussian regime, which has a comparable contribution to $\mathcal{P}_{\gamma\rightarrow\gamma}^{\textrm {tot}}$ as  ALP-photon oscillation inside the galaxies.

\section{Conclusions}

\label{sec:Conclusions}

We presented a comprehensive analytical framework for ALP-photon mixing in the presence of EBL attenuation, systematically treating constant, Gaussian stochastic, and non-Gaussian magnetic field configurations. Our work yields several key findings with direct implications for very-high-energy gamma-ray astrophysics:

\begin{itemize}
\item \textbf{For constant magnetic fields}, we derived exact closed-form probabilities and identified a perturbative plateau regime at $d \gg \lambda_\gamma$ where photon survival scales as $B^4$, isolating the four-point magnetic correlation as a sensitive probe of non-Gaussianity.

\item \textbf{For Gaussian stochastic magnetic fields}, we obtained—for the first time—analytical expressions for the non-exponential-decay components in the strong-attenuation regime. Contrary to the widely used domain-like model, the photon survival probability is suppressed by several orders of magnitude, while both conversion and survival probabilities exhibit distinct multi-peak structures arising from mass-equal resonance ($A=0$), stochastic resonance ($\lambda_B \lesssim  l_{\mathrm{osc}} $), and EBL attenuation.

\item \textbf{For non-Gaussian stochastic  magnetic fields}, we demonstrated that non-Gaussianity can enhance the photon survival probability by orders of magnitude relative to the Gaussian case—potentially explaining the unexpectedly high flux of $>10$ TeV photons from GRB 221009A observed by LHAASO  invoking ALPs.
\end{itemize}

Our results establish that stochastic magnetic fields cannot be reduced to domain-like coherence without losing essential physics, and that high-energy $\gamma$-ray spectra encode statistically accessible information about both the power spectrum and the non-Gaussian structure of intergalactic magnetic fields. The quartic scaling $\mathcal{P}_{\gamma\rightarrow\gamma} \sim B^4$ within the perturbative regime naturally connects photon survival to four-point magnetic correlations, offering a direct observational window into magnetic non-Gaussianity.

As next-generation observatories push toward PeV sensitivities, the effects identified here will become essential for distinguishing between astrophysical attenuation and new physics in the extragalactic very-high-energy sky. The distinctive spectral signatures predicted by our framework—peak structures from multiple resonances, non-exponential plateaus, and non-Gaussian enhancements—provide concrete targets for LHAASO, CTA, and future experiments to probe both ALP physics and the statistical properties of intergalactic magnetic fields.

\vspace{0.2cm}

\noindent\textbf{Acknowledgements. }
AA work is supported by the National Science Foundation of China (NSFC) through the grant No.\ 12350410358; the Talent Scientific Research Program of College of Physics, Sichuan University, Grant No.\ 1082204112427; the Fostering Program in Disciplines Possessing Novel Features for Natural Science of Sichuan University, Grant No.2020SCUNL209 and 1000 Talent program of Sichuan province 2021. YFC is supported in part by the National Key R\&D Program of China (2021YFC2203100), by NSFC (12433002), by CAS young interdisciplinary innovation team (JCTD-2022-20), by 111 Project (B23042), and by USTC Fellowship for International Cooperation.
SC, QG, and GL acknowledge the Istituto Nazionale di Fisica Nucleare (INFN), Sezione di Napoli, \textit{iniziative specifiche}   QGSKY and MOONLIGHT2.
This paper is based upon work from COST Action CA21136 {\it Addressing observational tensions in cosmology with systematics and fundamental physics} (CosmoVerse) supported by European Cooperation in Science and Technology.

\appendix

\section{}
%\subsection{Appendix A}
%\label{appA}

In this appendix section, we derive Eq. (\ref{eq:P-U}), which gives the relation
between the relevant probabilities in ALP-photon mixing and the evolution
operator. We begin with a brief review of the wavefunction, evolution
operator and density matrix \cite{Dolgov:2012be,Galanti:2018myb}.
Given a quantum state $\left|\Psi(t)\right\rangle $, its dynamics
satisfy the Schrodinger equation 
\begin{eqnarray}
i\frac{d}{dt}\left|\Psi(t)\right\rangle  & = & H(t)\left|\Psi(t)\right\rangle .\label{eq:waveEoM}
\end{eqnarray}
In terms of the evolution operator $U(t,t_{0})$ given by $\left|\Psi(t)\right\rangle =U(t,t_{0})\left|\Psi(t_{0})\right\rangle $,
its equation reads 
\begin{eqnarray}
i\frac{d}{dt}U(t,t_{0}) & = & H(t)U(t,t_{0}).\label{eq:UEoM}
\end{eqnarray}
Define the density matrix $\rho(t)=\left|\Psi(t)\right\rangle \left\langle \Psi(t)\right|$,
its relation to evolution operator is $\rho(t)=U(t,t_{0})\rho(t_{0})U^{\dagger}(t,t_{0})$.
It satisfies 
\begin{eqnarray}
i\frac{d}{dt}\rho(t) & = & H(t)\rho(t)-\rho(t)H^{\dagger}(t).\label{eq:densityEoM}
\end{eqnarray}
Note that EoM for the wavefunction (Eq. (\ref{eq:waveEoM})), the
evolution operator (Eq. (\ref{eq:UEoM})) and density matrix (Eq.
\ref{eq:densityEoM}) are equivalent. To illustrate the physical
meaning of density matrix, we consider the initial pure state $\left|\Psi(t_{0})\right\rangle =\left|\varphi\right\rangle $
with density matrix $\rho(t_{0})=\left|\varphi\right\rangle \left\langle \varphi\right|$.
Using Eq. (\ref{eq:densityEoM}), one obtains the density matrix $\rho(t)$
at arbitrary time $t$. The probability of detecting the sate $\left|\psi\right\rangle $
at $t$ is 
\begin{eqnarray}
\mathcal{P}(t)=\left\langle \psi\right|\rho(t)\left|\psi\right\rangle  & = & \textrm{Tr}\left(\left|\psi\right\rangle \left\langle \psi\right|\rho(t)\right),
\end{eqnarray}
where Tr is the trace operator. In other words, $\mathcal{P}(t)$
can be interpreted as the conversion probability from the initial
state $\left|\varphi\right\rangle $ to the final state $\left|\psi\right\rangle $. 

Applying this formalism to the APL-photon mixing, one can see that
the Schrodinger equation Eq. \ref{eq:waveEoM} is exactly same as
Eq. (\ref{eq:EoM-const}) by replacing time $t$ by propagation distance
$l$, and the Hamiltonian $H$ with kernel matrix $-K$. A general
quantum states can be written as 
\begin{eqnarray}
\left|\Psi(l)\right\rangle  & = & c_{1}(l)\left|A_{x}\right\rangle +c_{2}(l)\left|A_{y}\right\rangle +c_{3}(l)\left|a\right\rangle ,
\end{eqnarray}
where 
\begin{eqnarray}
\left|A_{x}\right\rangle =\left(\begin{array}{c}
1\\
0\\
0
\end{array}\right), &  & \left|A_{y}\right\rangle =\left(\begin{array}{c}
0\\
1\\
0
\end{array}\right),\left|a\right\rangle =\left(\begin{array}{c}
0\\
0\\
1
\end{array}\right)
\end{eqnarray}
form the basis representing the eigenstates of two photon polarization
states and the ALP state. 

Assuming the initial condition is a pure ALP state, $\left|\Psi(l_{0})\right\rangle =\left|a\right\rangle $, the corresponding density matrix is 
\begin{eqnarray}
\rho(l_{0})=\left|\Psi(l_{0})\right\rangle \left\langle \Psi(l_{0})\right| & = & \left(\begin{array}{ccc}
0\qquad & 0\qquad & 0\\
0\qquad & 0\qquad & 0\\
0\qquad & 0\qquad & 1
\end{array}\right).
\end{eqnarray}
One can either solve Eq. (\ref{eq:densityEoM}) to directly obtain
$\rho(l)$, or equivalently, first solve $U(t,t_{0})$ from Eq. (\ref{eq:UEoM})
and then compute $\rho(l)=U(l,l_{0})\rho(l_{0})U^{\dagger}(l,l_{0})$.
At distance $l$, the probability to obtain the states $\left|A_{x}\right\rangle $
and $\left|A_{y}\right\rangle $ is 
\begin{eqnarray}
\mathcal{P}_{a\rightarrow\gamma}(l) & = & \left\langle A_{x}\right|\rho(l)\left|A_{x}\right\rangle +\left\langle A_{y}\right|\rho(l)\left|A_{y}\right\rangle =\textrm{Tr}\left(\left|A_{x}\right\rangle \left\langle A_{x}\right|\rho(l)\right)+\textrm{Tr}\left(\left|A_{y}\right\rangle \left\langle A_{y}\right|\rho(l)\right)\nonumber \\
 & = & \textrm{Tr}\left(\left(\begin{array}{ccc}
1\qquad & 0\qquad & 0\\
0\qquad & 0\qquad & 0\\
0\qquad & 0\qquad & 0
\end{array}\right)U(l,l_{0})\rho(l_{0})U^{\dagger}(l,l_{0})\right)+\textrm{Tr}\left(\left(\begin{array}{ccc}
0\qquad & 0\qquad & 0\\
0\qquad & 1\qquad & 0\\
0\qquad & 0\qquad & 0
\end{array}\right)U(l,l_{0})\rho(l_{0})U^{\dagger}(l,l_{0})\right)\nonumber \\
 & = & U_{13}U_{13}^{*}+U_{23}U_{23}^{*},
\end{eqnarray}
where in last step we used the matrix form of $U$ and $\rho(l_{0})$
to take the trace operation. The probability $\mathcal{P}_{a\rightarrow\gamma}(l)$
can be interpreted as the conversion probability from ALP to photon
with two polarization modes after mixing system travels over a distance
$d=l-l_{0}$. Similarly, the ALP survival probability is
\begin{eqnarray}
\mathcal{P}_{a\rightarrow a} & = & Tr\left[\left|a\right\rangle \left\langle a\right|U(l,l_{0})\rho(l_{0})U^{\dagger}(l,l_{0})\right]=U_{33}U_{33}^{*}.
\end{eqnarray}

Now consider an initially unpolarized photon state with density matrix
given by
\begin{eqnarray}
\rho(l_{0}) & = & \frac{1}{2}\left|A_{x}\right\rangle \left\langle A_{x}\right|+\frac{1}{2}\left|A_{y}\right\rangle \left\langle A_{y}\right|=\frac{1}{2}\left(\begin{array}{ccc}
1\qquad & 0\qquad & 0\\
0\qquad & 1\qquad & 0\\
0\qquad & 0\qquad & 0
\end{array}\right),
\end{eqnarray}
using the same procedure, we obtain the photon survival probability
and photon-ALP conversion probability as
\begin{eqnarray}
\mathcal{P}_{\gamma\rightarrow\gamma} & = & \textrm{Tr}\left[\left|A_{x}\right\rangle \left\langle A_{x}\right|U(l,l_{0})\rho(l_{0})U^{\dagger}(l,l_{0})\right]+\textrm{Tr}\left[\left|A_{y}\right\rangle \left\langle A_{y}\right|U(l,l_{0})\rho(l_{0})U^{\dagger}(l,l_{0})\right]\nonumber \\
 & = & \frac{1}{2}\left(U_{11}U_{11}^{*}+U_{12}U_{12}^{*}+U_{21}U_{21}^{*}+U_{22}U_{22}^{*}\right),\nonumber \\
\mathcal{P}_{\gamma\rightarrow a} & = & \textrm{Tr}\left[\left|a\right\rangle \left\langle a\right|U(l,l_{0})\rho(l_{0})U^{\dagger}(l,l_{0})\right]\nonumber \\
 & = & \frac{1}{2}\left(U_{31}U_{31}^{*}+U_{32}U_{32}^{*}\right).
\end{eqnarray}

\section{}
%\subsection*{Appendix B}

In this appendix section, we derive the perturbation expansion of the relevant
probabilities with respect to the dimensionless parameter $g_{a\gamma}B\lambda_{\gamma}$.
An order-of-magnitude analysis is performed to determine the perturbative
regime. For ALP-photon mixing in the constant magnetic field, expanding
Eq. (\ref{eq:const-Ps}) yields 
\begin{eqnarray}
\mathcal{P}_{\gamma\leftrightarrow a} & \sim & \left(g_{a\gamma}B\lambda_{\gamma}\right)^{2}\frac{1}{1+A^{2}\lambda_{\gamma}^{2}}\left(1+e^{-\frac{d}{\lambda_{\gamma}}}+e^{-\frac{d}{2\lambda_{\gamma}}}\right)\nonumber \\
 &  & +\left(g_{a\gamma}B\lambda_{\gamma}\right)^{4}\frac{1}{\left(1+A^{2}\lambda_{\gamma}^{2}\right)^{3}}\left\{ \left(1+\frac{d}{\lambda_{\gamma}}\right)\left(1+A^{2}\lambda_{\gamma}^{2}\right)+e^{-\frac{d}{\lambda_{\gamma}}}+e^{-\frac{d}{2\lambda_{\gamma}}}\right\} \nonumber \\
 &  & +\left(g_{a\gamma}B\lambda_{\gamma}\right)^{6}\frac{1}{\left(1+A^{2}\lambda_{\gamma}^{2}\right)^{5}}\left\{ \left(1+\frac{d}{\lambda_{\gamma}}+\frac{d^{2}}{\lambda_{\gamma}^{2}}\right)\left(1+A^{2}\lambda_{\gamma}^{2}+A^{4}\lambda_{\gamma}^{4}\right)+e^{-\frac{d}{\lambda_{\gamma}}}+e^{-\frac{d}{2\lambda_{\gamma}}}\right\} \nonumber \\
 &  & +o\left(\left(g_{a\gamma}B\lambda_{\gamma}\right)^{8}\right),\label{eq:constB-P-a-ph-expan}
\end{eqnarray}
and 
\begin{eqnarray}
\mathcal{P}_{\gamma\rightarrow\gamma} & \sim & e^{-\frac{d}{\lambda_{\gamma}}}\nonumber \\
 &  & +\left(g_{a\gamma}B\lambda_{\gamma}\right)^{2}\frac{1}{\left(1+A^{2}\lambda_{\gamma}^{2}\right)^{2}}\left\{ e^{-\frac{d}{\lambda_{\gamma}}}+e^{-\frac{d}{2\lambda_{\gamma}}}\right\} \nonumber \\
 &  & +\left(g_{a\gamma}B\lambda_{\gamma}\right)^{4}\frac{1}{\left(1+A^{2}\lambda_{\gamma}^{2}\right)^{4}}\left\{ \left(1+A^{2}\lambda_{\gamma}^{2}+A^{4}\lambda_{\gamma}^{4}\right)+e^{-\frac{d}{\lambda_{\gamma}}}+e^{-\frac{d}{2\lambda_{\gamma}}}\right\} \nonumber \\
 &  & +\left(g_{a\gamma}B\lambda_{\gamma}\right)^{6}\frac{1}{\left(1+A^{2}\lambda_{\gamma}^{2}\right)^{6}}\left\{ \left(1+\frac{d}{\lambda_{\gamma}}\right)\left(1+A^{2}\lambda_{\gamma}^{2}+A^{4}\lambda_{\gamma}^{4}+A^{6}\lambda_{\gamma}^{6}\right)+e^{-\frac{d}{\lambda_{\gamma}}}+e^{-\frac{d}{2\lambda_{\gamma}}}\right\} \nonumber \\
 &  & +\left(g_{a\gamma}B\lambda_{\gamma}\right)^{8}\frac{1}{\left(1+A^{2}\lambda_{\gamma}^{2}\right)^{8}}\left\{ \left(1+\frac{d}{\lambda_{\gamma}}+\frac{d^{2}}{\lambda_{\gamma}^{2}}\right)\left(1+A^{2}\lambda_{\gamma}^{2}+A^{4}\lambda_{\gamma}^{4}+A^{6}\lambda_{\gamma}^{6}+A^{8}\lambda_{\gamma}^{8}\right)+e^{-\frac{d}{\lambda_{\gamma}}}+e^{-\frac{d}{2\lambda_{\gamma}}}\right\} \nonumber \\
 &  & +o\left(\left(g_{a\gamma}B\lambda_{\gamma}\right)^{10}\right),\label{eq:constB-ph-su-expan}
\end{eqnarray}
and 
\begin{eqnarray}
\mathcal{P}_{a\rightarrow a} & \sim & 1+\left(g_{a\gamma}B\lambda_{\gamma}\right)^{2}\frac{1}{\left(1+A^{2}\lambda_{\gamma}^{2}\right)^{2}}\left\{ \left(1+\frac{d}{\lambda_{\gamma}}\right)\left(1+A^{2}\lambda_{\gamma}^{2}\right)+e^{-\frac{d}{2\lambda_{\gamma}}}\right\} +o\left(\left(g_{a\gamma}B\lambda_{\gamma}\right)^{4}\right).\label{eq:constB-a-su-expan}
\end{eqnarray}
In the strong EBL attenuation regime, $d\gg\lambda_{\gamma}$, the
exponential-decay term is much suppressed, then we obtain
\begin{eqnarray}
\mathcal{P}_{\gamma\leftrightarrow a} & \sim & \begin{cases}
\left(g_{a\gamma}B\lambda_{\gamma}\right)^{2}+\left(g_{a\gamma}B\lambda_{\gamma}\right)^{4}\frac{d}{\lambda_{\gamma}}+\left(g_{a\gamma}B\lambda_{\gamma}\right)^{6}\frac{d^{2}}{\lambda_{\gamma}^{2}}+... & A\lambda_{\gamma}\lesssim\mathcal{O}(1),\\
\left(g_{a\gamma}B\lambda_{\gamma}\right)^{2}\frac{1}{\left(A\lambda_{\gamma}\right)^{2}}+\left(g_{a\gamma}B\lambda_{\gamma}\right)^{4}\frac{d}{\lambda_{\gamma}}\frac{1}{\left(A\lambda_{\gamma}\right)^{4}}+\left(g_{a\gamma}B\lambda_{\gamma}\right)^{6}\frac{d^{2}}{\lambda_{\gamma}^{2}}\frac{1}{\left(A\lambda_{\gamma}\right)^{6}}+... & A\lambda_{\gamma}\gg1,
\end{cases}\nonumber \\
\mathcal{P}_{\gamma\rightarrow\gamma} & \sim & \begin{cases}
\left(g_{a\gamma}B\lambda_{\gamma}\right)^{4}+\left(g_{a\gamma}B\lambda_{\gamma}\right)^{6}\frac{d}{\lambda_{\gamma}}+\left(g_{a\gamma}B\lambda_{\gamma}\right)^{8}\frac{d^{2}}{\lambda_{\gamma}^{2}}+... & A\lambda_{\gamma}\lesssim\mathcal{O}(1),\\
\left(g_{a\gamma}B\lambda_{\gamma}\right)^{4}\frac{1}{\left(A\lambda_{\gamma}\right)^{4}}+\left(g_{a\gamma}B\lambda_{\gamma}\right)^{6}\frac{d}{\lambda_{\gamma}}\frac{1}{\left(A\lambda_{\gamma}\right)^{6}}+\left(g_{a\gamma}B\lambda_{\gamma}\right)^{8}\frac{d^{2}}{\lambda_{\gamma}^{2}}\frac{1}{\left(A\lambda_{\gamma}\right)^{8}}+... & A\lambda_{\gamma}\gg1,
\end{cases}\nonumber \\
\mathcal{P}_{a\rightarrow a} & \sim & \begin{cases}
1+\left(g_{a\gamma}B\lambda_{\gamma}\right)^{2}\frac{d}{\lambda_{\gamma}}+... & A\lambda_{\gamma}\lesssim\mathcal{O}(1),\\
1+\left(g_{a\gamma}B\lambda_{\gamma}\right)^{2}\frac{d}{\lambda_{\gamma}}\frac{1}{\left(A\lambda_{\gamma}\right)^{2}}+... & A\lambda_{\gamma}\gg1.
\end{cases}\label{eq:appen-P-const-expansion}
\end{eqnarray}
It shows that the perturbation condition is $\left(g_{a\gamma}B\lambda_{\gamma}\right)^{2}\frac{d}{\lambda_{\gamma}}\ll1$
for $A\lambda_{\gamma}\lesssim\mathcal{O}(1)$, and $\left(g_{a\gamma}B\lambda_{\gamma}\right)^{2}\frac{d}{\lambda_{\gamma}}\frac{1}{\left(A\lambda_{\gamma}\right)^{2}}\ll1$
for $A\lambda_{\gamma}\gg1$. For order-of-magnitude estimates, these
two region $A\lambda_{\gamma}\lesssim\mathcal{O}(1)$ and $A\lambda_{\gamma}\gg1$
shown above can be combined into a compact expression given in Eq.
(\ref{eq:P-expansion}). 

In the short distance $d<\lambda_{\gamma}$, the EBL absorption can
be neglected, and Eq. (\ref{eq:constB-P-a-ph-expan}) reduces to 
\begin{eqnarray}
\mathcal{P}_{\gamma\leftrightarrow a} & \sim & \begin{cases}
\left(g_{a\gamma}B\lambda_{\gamma}\right)^{2}+\left(g_{a\gamma}B\lambda_{\gamma}\right)^{4}+\left(g_{a\gamma}B\lambda_{\gamma}\right)^{6}+... & A\lambda_{\gamma}\lesssim\mathcal{O}(1),\\
\left(g_{a\gamma}B\lambda_{\gamma}\right)^{2}\frac{1}{\left(A\lambda_{\gamma}\right)^{2}}+\left(g_{a\gamma}B\lambda_{\gamma}\right)^{4}\frac{1}{\left(A\lambda_{\gamma}\right)^{4}}+\left(g_{a\gamma}B\lambda_{\gamma}\right)^{6}\frac{1}{\left(A\lambda_{\gamma}\right)^{6}}+... & A\lambda_{\gamma}\gg1.
\end{cases}
\end{eqnarray}
The perturbation condition becomes $\left(g_{a\gamma}B\lambda_{\gamma}\right)^{2}\ll1$
for $A\lambda_{\gamma}\lesssim\mathcal{O}(1)$, and $\left(g_{a\gamma}B\lambda_{\gamma}\right)^{2}\frac{1}{\left(A\lambda_{\gamma}\right)^{2}}\ll1$
for $A\lambda_{\gamma}\gg1$. For sufficient weak EBL attenuation
where $\lambda_{\gamma}$ is large enough at $A\lambda_{\gamma}\gg1$,
the perturbation condition $g_{a\gamma}Bl_{osc}\ll1$ ($l_{osc}\sim|A|^{-1}$)
corresponds to weak mixing angle of ALP-photon oscillation \cite{raffelt1988mixing}.
Regarding the photon survival probability, the unitary conservation
in weak EBL regime requires $\mathcal{P}_{\gamma\rightarrow a}+\mathcal{P}_{\gamma\rightarrow\gamma}\simeq1$.
Therefore, the leading order of photon survival probability is around
unity for $d<\lambda_{\gamma}$. 

A similar analysis applies to stochastic magnetic field. Although
the general treatment involves complicate integrals over $\int d^{3}k$
and $\int dl$, we adopt two simplifications that would not alter
the main conclusion. First, we study the ALP-photon mixing in two
spatial dimensions instead of three. Second, we focus on the monochromatic
spectrum of magnetic field, with correlation function $\left\langle B(l)B(l')\right\rangle \sim e^{ik_{B}(l-l')}$,
which allows the magnetic field to be treated effectively as an oscillatory
field $B(l)\sim Be^{ik_{B}l}$.Note that here \(B(l)\) is treated as a complex field; consequently, the complex conjugate \(B^{*}(l)\) must be included in the evolution operator where appropriate. Within
these simplifications, from Eq. (\ref{eq:U-expan}) we obtain {\small
\begin{eqnarray}
U_{12} & \sim & g_{a\gamma}e^{-bd}\int_{0}^{d}dl_{1}e^{l_{1}b}e^{-il_{1}A}B(l_{1})\nonumber \\
 &  & +g_{a\gamma}^{3}e^{-bd}\int_{0}^{d}dl_{1}\int_{0}^{l_{1}}dl_{2}\int_{0}^{l_{2}}dl_{3}e^{\left(l_{1}-l_{2}+l_{3}\right)b}e^{-i\left(l_{1}-l_{2}+l_{3}\right)A}B(l_{1})B^{*}(l_{2})B(l_{3})\nonumber \\
 &  & +g_{a\gamma}^{5}e^{-bd}\int_{0}^{d}dl_{1}\int_{0}^{l_{1}}dl_{2}\int_{0}^{l_{2}}dl_{3}\int_{0}^{l_{3}}dl_{4}\int_{0}^{l_{4}}dl_{5}e^{\left(l_{1}-l_{2}+l_{3}-l_{4}+l_{5}\right)b}e^{-i\left(l_{1}-l_{2}+l_{3}-l_{4}+l_{5}\right)A}B(l_{1})B^{*}(l_{2})B(l_{3})B^{*}(l_{4})B(l_{5})\nonumber \\
 &  & +...\nonumber \\
 & \sim & g_{a\gamma}Be^{-bd}\int_{0}^{d}dl_{1}e^{l_{1}b}e^{-il_{1}(A-k_{B})}\nonumber \\
 &  & +g_{a\gamma}^{3}B^{3}e^{-bd}\int_{0}^{d}dl_{1}\int_{0}^{l_{1}}dl_{2}\int_{0}^{l_{2}}dl_{3}e^{\left(l_{1}-l_{2}+l_{3}\right)b}e^{-i\left(l_{1}-l_{2}+l_{3}\right)(A-k_{B})}\nonumber \\
 &  & +g_{a\gamma}^{5}B^{5}e^{-bd}\int_{0}^{d}dl_{1}\int_{0}^{l_{1}}dl_{2}\int_{0}^{l_{2}}dl_{3}\int_{0}^{l_{3}}dl_{4}\int_{0}^{l_{4}}dl_{5}e^{\left(l_{1}-l_{2}+l_{3}-l_{4}+l_{5}\right)b}e^{-i\left(l_{1}-l_{2}+l_{3}-l_{4}+l_{5}\right)(A-k_{B})}\nonumber \\
 &  & +...,
\end{eqnarray}
}and {\small
\begin{eqnarray}
U_{11} & \sim & e^{-bd}\nonumber \\
 &  & +g_{a\gamma}^{2}B^{2}e^{-bd}\int_{0}^{d}dl_{1}\int_{0}^{l_{1}}dl_{2}e^{(l_{1}-l_{2})b}e^{-i\left(l_{1}-l_{2}\right)(A-k_{B})}\nonumber \\
 &  & +g_{a\gamma}^{4}B^{4}e^{-bd}\int_{0}^{d}dl_{1}\int_{0}^{l_{1}}dl_{2}\int_{0}^{l_{2}}dl_{3}\int_{0}^{l_{3}}dl_{4}e^{\left(l_{1}-l_{2}+l_{3}-l_{4}\right)b}e^{-i\left(l_{1}-l_{2}+l_{3}-l_{4}\right)(A-k_{B})}\nonumber \\
 &  & +g_{a\gamma}^{6}B^{6}e^{-bd}\int_{0}^{d}dl_{1}\int_{0}^{l_{1}}dl_{2}\int_{0}^{l_{2}}dl_{3}\int_{0}^{l_{3}}dl_{4}\int_{0}^{l_{4}}dl_{5}\int_{0}^{l_{5}}dl_{6}e^{\left(l_{1}-l_{2}+l_{3}-l_{4}+l_{5}-l_{6}\right)b}e^{-i\left(l_{1}-l_{2}+l_{3}-l_{4}+l_{5}-l_{6}\right)(A-k_{B})}\nonumber \\
 &  & +g_{a\gamma}^{8}B^{8}e^{-bd}\int_{0}^{d}dl_{1}\int_{0}^{l_{1}}dl_{2}\int_{0}^{l_{2}}dl_{3}\int_{0}^{l_{3}}dl_{4}\int_{0}^{l_{4}}dl_{5}\int_{0}^{l_{5}}dl_{6}\int_{0}^{l_{6}}dl_{7}\int_{0}^{l_{7}}dl_{8}\nonumber \\
 &  & \quad\times e^{\left(l_{1}-l_{2}+l_{3}-l_{4}+l_{5}-l_{6}+l_{7}-l_{8}\right)b}e^{-i\left(l_{1}-l_{2}+l_{3}-l_{4}+l_{5}-l_{6}+l_{7}-l_{8}\right)(A-k_{B})}\nonumber \\
 &  & +....
\end{eqnarray}
}In the strong EBL region $d\gg\lambda_{\gamma}$, we drop the exponential-decay
terms and obtain
\begin{eqnarray}
\mathcal{P}_{a\rightarrow\gamma} & \sim & |U_{12}|^{2}\sim\begin{cases}
\left(g_{a\gamma}B\lambda_{\gamma}\right)^{2}+\left(g_{a\gamma}B\lambda_{\gamma}\right)^{4}\frac{d}{\lambda_{\gamma}}+\left(g_{a\gamma}B\lambda_{\gamma}\right)^{6}\frac{d^{2}}{\lambda_{\gamma}^{2}}+... & \widetilde{A}\lambda_{\gamma}\lesssim\mathcal{O}(1),\\
\left(g_{a\gamma}B\lambda_{\gamma}\right)^{2}\frac{1}{\left(\widetilde{A}\lambda_{\gamma}\right)^{2}}+\left(g_{a\gamma}B\lambda_{\gamma}\right)^{4}\frac{d}{\lambda_{\gamma}}\frac{1}{\left(\widetilde{A}\lambda_{\gamma}\right)^{4}}+\left(g_{a\gamma}B\lambda_{\gamma}\right)^{6}\frac{d^{2}}{\lambda_{\gamma}^{2}}\frac{1}{\left(\widetilde{A}\lambda_{\gamma}\right)^{6}}+... & \widetilde{A}\lambda_{\gamma}\gg1,
\end{cases}
\end{eqnarray}
and 
\begin{eqnarray}
\mathcal{P}_{\gamma\rightarrow\gamma} & \sim & |U_{11}|^{2}\sim\begin{cases}
\left(g_{a\gamma}B\lambda_{\gamma}\right)^{4}+\left(g_{a\gamma}B\lambda_{\gamma}\right)^{6}\frac{d}{\lambda_{\gamma}}+\left(g_{a\gamma}B\lambda_{\gamma}\right)^{8}\frac{d^{2}}{\lambda_{\gamma}^{2}}+... & \widetilde{A}\lambda_{\gamma}\lesssim\mathcal{O}(1),\\
\left(g_{a\gamma}B\lambda_{\gamma}\right)^{4}\frac{1}{\left(\widetilde{A}\lambda_{\gamma}\right)^{4}}+\left(g_{a\gamma}B\lambda_{\gamma}\right)^{6}\frac{d}{\lambda_{\gamma}}\frac{1}{\left(\widetilde{A}\lambda_{\gamma}\right)^{6}}+\left(g_{a\gamma}B\lambda_{\gamma}\right)^{8}\frac{d^{2}}{\lambda_{\gamma}^{2}}\frac{1}{\left(\widetilde{A}\lambda_{\gamma}\right)^{8}}+... & \widetilde{A}\lambda_{\gamma}\gg1,
\end{cases}\label{eq:appen-P-phph-expan}
\end{eqnarray}
where $\widetilde{A}\sim A-k_{B}$. Similarly to the constant magnetic
field case (Eq. (\ref{eq:appen-P-const-expansion})), the perturbative
condition is $\left(g_{a\gamma}B\lambda_{\gamma}\right)^{2}\frac{d}{\lambda_{\gamma}}\frac{1}{1+\left(\widetilde{A}\lambda_{\gamma}\right)^{2}}\ll1$
in the strong EBL attenuation regime $d\gg\lambda_{\gamma}$. For
typical intergalactic parameters, e.g. $g_{a\gamma}=10^{-12}\textrm{GeV}^{-1}$,
$B=10^{-9}\textrm{Gs}$, $d=\textrm{Gpc}$ and $\omega\gtrsim20$TeV,
this conservatively satisfies $\left(g_{a\gamma}B\lambda_{\gamma}\right)^{2}\frac{d}{\lambda_{\gamma}}\lesssim1$.
Reducing \(g_{a\gamma}\) or \(B\) by an order of magnitude ensures that the perturbative regime is well satisfied, extending the valid energy range down to \(\omega \sim 0.1\) TeV. 
Moreover, the leading contribution
to $\mathcal{P}_{\gamma\rightarrow\gamma}$ in Eq. (\ref{eq:appen-P-phph-expan}),
$\left(g_{a\gamma}B\lambda_{\gamma}\right)^{4}/\left(1+\lambda_{\gamma}^{4}\left(A-k_{B}\right)^{4}\right)$
(valid for all $\widetilde{A}\lambda_{\gamma}$), partially justifies
the parameterized expression of $\mathcal{P}_{\gamma\rightarrow\gamma}$
given in Eq. (\ref{eq:P-parameterization}). 

\section{}
%\subsection*{Appendix C}

In this appendix section, we present the complete expression of several formula
quoted in the main text. The full expression of $\mathcal{P}_{a\rightarrow\gamma}$
in the Gaussian magnetic field case is (Eq. (\ref{eq:stoB-P-a-ph})
){\small
\begin{eqnarray}
\mathcal{P}_{a\rightarrow\gamma} & = & \frac{1}{4}g_{a\gamma}^{2}\frac{1}{(2\pi)^{3}}\int dk2\pi k^{2}P_{B}\int_{-1}^{1}dt\left(1+t^{2}\right)\frac{1+e^{-2bd}-e^{-bd}2\textrm{cos}\left(\left(A+kt\right)d\right)}{\left(A+kt\right)^{2}+b^{2}}\nonumber \\
 & = & \frac{1}{4}g_{a\gamma}^{2}\frac{1}{(2\pi)^{3}}\int dk2\pi k^{2}P_{B}\frac{1}{bk^{3}}\nonumber \\
 &  & \times\left(\left(A^{2}-b^{2}+k^{2}\right)\left(\textrm{arctan}\left(\frac{A+k}{b}\right)-\textrm{arctan}\left(\frac{A-k}{b}\right)\right)+b\left(2k+A\textrm{ln}\left(\frac{(A-k)^{2}+b^{2}}{(A+k)^{2}+b^{2}}\right)\right)\right)\nonumber \\
 &  & +\frac{1}{4}g_{a\gamma}^{2}e^{-2bd}\frac{1}{(2\pi)^{3}}\int dk2\pi k^{2}P_{B}\frac{1}{bk^{3}}\nonumber \\
 &  & \times\left(\left(A^{2}-b^{2}+k^{2}\right)\left(\textrm{arctan}\left(\frac{A+k}{b}\right)-\textrm{arctan}\left(\frac{A-k}{b}\right)\right)+b\left(2k+A\textrm{ln}\left(\frac{(A-k)^{2}+b^{2}}{(A+k)^{2}+b^{2}}\right)\right)\right)\nonumber \\
 &  & +\frac{1}{4}g_{a\gamma}^{2}e^{-db}\frac{1}{(2\pi)^{3}}\int2\pi k^{2}P_{B}dk\frac{2}{bk^{3}}\left\{ \frac{b}{d}\textrm{sin}\left(d\left(A-k\right)\right)-\frac{b}{d}\textrm{sin}\left(d\left(A+k\right)\right)\right.\nonumber \\
 &  & +\left(A^{2}-b^{2}+k^{2}\right)\textrm{cosh}\left(db\right)\textrm{Im}\left(\textrm{Ci}\left(d(A-ib-k)\right)\right)-2Ab\textrm{cosh}\left(db\right)\textrm{Re}\left(\textrm{Ci}\left(d(A-ib-k)\right)\right)\nonumber \\
 &  & -\left(A^{2}-b^{2}+k^{2}\right)\textrm{cosh}\left(db\right)\textrm{Im}\left(\textrm{Ci}\left(d(A-ib+k)\right)\right)+2Ab\textrm{cosh}\left(db\right)\textrm{Re}\left(\textrm{Ci}\left(d(A-ib+k)\right)\right)\nonumber \\
 &  & -\left(A^{2}-b^{2}+k^{2}\right)\textrm{sinh}\left(db\right)\textrm{Re}\left(\textrm{Si}\left(d(A-ib-k)\right)\right)-2Ab\textrm{sinh}\left(db\right)\textrm{Im}\left(\textrm{Si}\left(d(A-ib-k)\right)\right)\nonumber \\
 &  & \left.+\left(A^{2}-b^{2}+k^{2}\right)\textrm{sinh}\left(db\right)\textrm{Re}\left(\textrm{Si}\left(d(A-ib+k)\right)\right)+2Ab\textrm{sinh}\left(db\right)\textrm{Im}\left(\textrm{Si}\left(d(A-ib+k)\right)\right)\right\} ,\label{eq:appen-P-aph-sto}
\end{eqnarray}
}where $\textrm{Si}(z)=\int_{0}^{z}\textrm{sin}(t)/tdt$, $\textrm{Ci}(z)=-\int_{z}^{\infty}\textrm{cos}(t)/tdt$, $\textrm{arctan}(z)$ is the inverse tangent function, $\textrm{Re}$ is the real part and $\textrm{Im}$ is the imaginary part. Here, the
first integral term corresponds to the non exponential-decay part,
which dominates in the strong EBL attenuation regime. 

The expression of the non exponential-decay part of $\mathcal{P}_{\gamma\rightarrow\gamma}$
in the Gaussian magnetic field is (Eq. \ref{eq:stoB-P-ph-ph}){\small
\begin{eqnarray}
\mathcal{P}_{\gamma\rightarrow\gamma} & = & \frac{1}{16}g_{a\gamma}^{4}\frac{\pi^{2}}{(2\pi)^{6}}\int dkk^{2}P_{B}(k)\int dk'k'{}^{2}P_{B}(k')\int_{-1}^{1}dt\left(1+t^{2}\right)\int_{-1}^{1}dt'\left(1+t'{}^{2}\right)\frac{e^{id(kt+k't')}}{\left(A-ib+kt\right)^{2}\left(A+ib-k't'\right)^{2}}\nonumber \\
 & = & \frac{1}{16}g_{a\gamma}^{4}\frac{\pi^{2}}{(2\pi)^{6}}\int dkk^{2}P_{B}(k)\int dk'k'{}^{2}P_{B}(k')\frac{1}{d^{2}k^{3}k'{}^{3}}\nonumber \\
 &  & \times\left\{ \frac{e^{-ikd}\left(-A\left(1-2bd\right)+k+i\left(b+\left(A^{2}-b^{2}+k^{2}\right)d\right)\right)}{A-k-ib}+\frac{e^{ikd}\left(A\left(1-2bd\right)+k-i\left(b+\left(A^{2}-b^{2}+k^{2}\right)d\right)\right)}{A+k-ib}\right.\nonumber \\
 &  & \left.+de^{-iAd}\left(2b+\left(A^{2}-b^{2}+k^{2}\right)d+i2A(1-bd)\right)e^{-bd}\left(\textrm{Ei}\left[\left(i\left(A-k\right)+b\right)d\right]-\textrm{Ei}\left[\left(i\left(A+k\right)+b\right)d\right]\right)\right\} \nonumber \\
 &  & \times\left\{ \frac{e^{-ik'd}\left(A\left(1-2bd\right)+k'+i\left(b+\left(A^{2}-b^{2}+k'{}^{2}\right)d\right)\right)}{A+k'+ib}+\frac{e^{ik'd}\left(A\left(1-2bd\right)-k'+i\left(b+\left(A^{2}-b^{2}+k'{}^{2}\right)d\right)\right)}{-A+k'-ib}\right.\nonumber \\
 &  & \left.+de^{iAd}\left(2b+\left(A^{2}-b^{2}+k'{}^{2}\right)d-i2A(1-bd)\right)e^{-bd}\left(\textrm{Ei}\left[-\left(i\left(A-k'\right)-b\right)d\right]-\textrm{Ei}\left[-\left(i\left(A+k'\right)-b\right)d\right]\right)\right\} ,\label{eq:appen-P-sto-phph}
\end{eqnarray}
}where $\textrm{Ei}(z)=-\int_{-z}^{\infty}e^{-t}/tdt$.

The expression of the non exponential-decay part of $\mathcal{P}_{\gamma\rightarrow\gamma}$
in the stochastic magnetic field with non-Gaussian case 2 is (Eq. (\ref{eq:NGcase2-P}))
{\small
\begin{eqnarray}
\mathcal{P}_{\gamma\rightarrow\gamma}^{\textrm{NG}} & = & 4\frac{1}{2}\frac{1}{16}g_{a\gamma}^{4}2\pi\int k^{2}dkP_{\rho}(k)\int_{-1}^{1}dt\frac{-e^{idkt}(1-2bd-idkt)}{(2b+ikt)^{4}}\nonumber \\
 & = & 4\frac{1}{2}\frac{1}{16}g_{a\gamma}^{4}2\pi\int k^{2}dkP_{\rho}(k)\left\{ e^{-2bd}\frac{d^{3}}{3k}i\left(\textrm{Ei}(2bd-idk)-\textrm{Ei}(2bd+idk)\right)\right.\nonumber \\
 &  & +2\frac{(-12b^{2}+16b^{3}d+16b^{4}d^{2}+k^{2}+4bdk^{2}+8b^{2}d^{2}k^{2}+d^{2}k^{4})}{3(4b^{2}+k^{2})^{3}}\textrm{cos}(dk)\nonumber \\
 &  & \left.+2\frac{(8b^{3}-16b^{4}d-32b^{5}d^{2}-6bk^{2}-16b^{3}d^{2}k^{2}+dk^{4}-2bd^{2}k^{4})}{3k(4b^{2}+k^{2})^{3}}\textrm{sin}(dk)\right\} \label{eq:appen-NG2}
\end{eqnarray}
}{\small\par}

The expression of the non exponential-decay part of $\mathcal{P}_{\gamma\rightarrow\gamma}$
in the stochastic magnetic field with non-Gaussian case 3 is (Eq.
(\ref{eq:NGcase3-P})){\small
\begin{eqnarray}
\mathcal{P}_{\gamma\rightarrow\gamma}^{\textrm{NG}} & = & \frac{1}{2^{9}\pi^{2}}g_{a\gamma}^{4}B_{rms}^{4}\frac{54}{25}g_{NL}\int_{-1}^{+1}dt\left(\frac{1}{A^{4}+2A^{2}(b^{2}-k_{B}^{2}t^{2})+(b^{2}+k_{B}^{2}t^{2})^{2}}+\frac{e^{2ik_{B}td}}{(A^{2}+(b+ik_{B}t)^{2})^{2}}+\frac{1}{(b^{2}+(A+k_{B}t)^{2})^{2}}\right)\nonumber \\
 & = & \frac{1}{2^{9}\pi^{2}}g_{a\gamma}^{4}B_{rms}^{4}\frac{54}{25}g_{NL}\left\{ \frac{1}{2Ab(A^{2}+b^{2})k_{B}}\left((iA+b)\textrm{arctanh}\left(\frac{k_{B}}{A+ib}\right)-(iA-b)\textrm{arctanh}\left(\frac{k_{B}}{A-ib}\right)\right)\right.\nonumber \\
 &  & -\frac{k_{B}(-A^{2}+b^{2}+k_{B}^{2})\textrm{cos}(2dk_{B})-b(A^{2}+b^{2}+k_{B}^{2})\textrm{sin}(2dk_{B})}{A^{2}k_{B}(A^{2}+b^{2}-2Ak_{B}+k_{B}^{2})(A^{2}+b^{2}+2Ak_{B}+k_{B}^{2})}\nonumber \\
 &  & +\frac{(-1+2iAd)e^{-2bd}}{4A^{3}k_{B}}e^{2iAd}\textrm{Ei}\left(-2id(A+ib-k_{B})\right)+\frac{(-1-2iAd)e^{-2bd}}{4A^{3}k_{B}}e^{-2iAd}\textrm{Ei}\left(2id(A-ib-k_{B})\right)\nonumber \\
 &  & +\frac{(1-2iAd)e^{-2bd}}{4A^{3}k_{B}}e^{2iAd}\textrm{Ei}\left(-2id(A+ib+k_{B})\right)+\frac{(1+2iAd)e^{-2bd}}{4A^{3}k_{B}}e^{-2iAd}\textrm{Ei}\left(2id(A-ib+k_{B})\right)\nonumber \\
 &  & \left.-\frac{1}{2b^{3}k_{B}}\left(\frac{b(A-k_{B})}{b^{2}+(A-k_{B})^{2}}-\frac{b(A+k_{B})}{b^{2}+(A+k_{B})^{2}}+\textrm{arctan}\left(\frac{A-k_{B}}{b}\right)-\textrm{arctan}\left(\frac{A+k_{B}}{b}\right)\right)\right\} ,\label{eq:Appen-NGcase3}
\end{eqnarray}
where }$\textrm{arctanh}(z)$ is the inverse function of $\textrm{tanh}(z)$. 

\bibliographystyle{unsrturl}

\end{document}